\documentclass[superscriptaddress,amsmath,amssymb,aps,prx,longbibliography,10pt]{revtex4-2}
\usepackage{preamble}
\usepackage{svg}
\usepackage{lineno}

\begin{document}
\title{Trion gas on the surface of a failed excitonic insulator.}


\author{Yuval Nitzav}
\affiliation{Department of Physics, Technion, Haifa, 3200003, Israel}
\author{Abigail Dishi}
\affiliation{Department of Physics, Technion, Haifa, 3200003, Israel}
\author{Himanshu Lohani}
\affiliation{Department of Physics, Technion, Haifa, 3200003, Israel}
\author{Ittai Sidilkover}
\affiliation{School of Physics and Astronomy, Faculty of Exact Sciences, Tel Aviv University, Tel-Aviv 6997801, Israel}
\affiliation{Center for Light-Matter Interaction, Tel Aviv University, Tel Aviv 6997801, Israel}
\author{Noam Ophir}
\affiliation{Department of Physics, Technion, Haifa, 3200003, Israel}
\author{Roni Anna Gofman}
\affiliation{Department of Physics, Technion, Haifa, 3200003, Israel}
\author{Avior Almoalem}
\affiliation{Department of Physics, Technion, Haifa, 3200003, Israel}
\author{Ilay Mangel}
\affiliation{Department of Physics, Technion, Haifa, 3200003, Israel}
 
\author{Nitzan Ragoler}
\affiliation{Department of Physics, Technion, Haifa, 3200003, Israel}

\author{Francois Bertran}
\affiliation{SOLEIL Synchrotron, L’Orme des Merisiers, Départementale 128, 91190 Saint-Aubin, France}
\author{Jaime Sánchez-Barriga}
\affiliation{Helmholtz-Zentrum Berlin für Materialien und Energie, Elektronenspeicherring BESSY II,
Albert-Einstein-Strasse 15, 12489 Berlin, Germany}
\affiliation{ IMDEA Nanoscience, C/ Faraday 9, Campus de Cantoblanco, 28049 Madrid, Spain}
\author{Dmitry Marchenko}
\affiliation{Helmholtz-Zentrum Berlin für Materialien und Energie, Elektronenspeicherring BESSY II,
Albert-Einstein-Strasse 15, 12489 Berlin, Germany}
\author{Andrei Varykhalov}
\affiliation{Helmholtz-Zentrum Berlin für Materialien und Energie, Elektronenspeicherring BESSY II,
Albert-Einstein-Strasse 15, 12489 Berlin, Germany}
\author{Nicholas Clark Plumb}
\affiliation{Photon Science Division, Paul Scherrer Institute, CH-5232 Villigen PSI, Switzerland}
\author{Irena Feldman}
\affiliation{Department of Physics, Technion, Haifa, 3200003, Israel}
\author{Hadas Soifer}
\affiliation{School of Physics and Astronomy, Faculty of Exact Sciences, Tel Aviv University, Tel-Aviv 6997801, Israel}
\affiliation{Center for Light-Matter Interaction, Tel Aviv University, Tel Aviv 6997801, Israel}
\author{Anna Keselman}
\affiliation{Department of Physics, Technion, Haifa, 3200003, Israel}
\affiliation{The Helen Diller Quantum Center, Technion, Haifa, 3200003, Israel}
\author{Amit Kanigel}
\affiliation{Department of Physics, Technion, Haifa, 3200003, Israel}
\affiliation{The Helen Diller Quantum Center, Technion, Haifa, 3200003, Israel}


\begin{abstract}
Trions, three-body bound states composed of an exciton and an additional charge, are typically fragile and require external excitation to form. Here, we report the spontaneous emergence of a stable trion gas at the surface of the layered semiconductor Ta$_2$NiS$_5$, revealed through angle-resolved photoemission spectroscopy. We observe a sharp, highly localized in-gap feature that cannot be explained by conventional band-theory. Instead, we argue that it arises from the formation of negative trions, stabilized by surface-induced band bending and the material’s quasi-one-dimensional geometry. Unlike excitons, these trions form without optical pumping and persist at equilibrium, marking a rare example of an interaction-driven surface state in a nominally conventional semiconductor. Our findings establish Ta$_2$NiS$_5$ as a unique platform for exploring many-body physics at surfaces and open new avenues for studying and controlling collective excitations in low-dimensional systems.

\end{abstract}
\maketitle

\section{ Introduction}
Quasiparticles composed of multiple interacting particles underpin much of modern condensed matter physics. While single particle excitations such as electrons and holes are ubiquitous, composite bound states such as excitons, polarons and Cooper pairs emerge only under special conditions and can drive new phases of matter. Among the most intriguing are trions, three body bound states of two electrons and one hole (or vice versa), which are a hallmark of strong Coulomb interactions in reduced dimensions. In semiconductors and two dimensional materials, trions are usually generated transiently by optical excitation and decay within picoseconds, making equilibrium trions in solid state systems without continuous pumping a long standing challenge.

 Excitons govern light absorption and emission in many semiconductors and become especially prominent when quantum confinement raises their binding energy above thermal fluctuations. In doped systems or in the presence of excess carriers, an exciton can bind an additional electron or hole to form a trion, or charged exciton, producing distinct low energy optical features \cite{lampert1958mobile}.
 
Two dimensional materials such as transition metal dichalcogenides (TMDs) provide an ideal setting, where strong Coulomb interactions and reduced dielectric screening yield large exciton and trion binding energies, enabling direct observation in photoluminescence and absorption experiments \cite{mak2013tightly,wang2018colloquium}.

Excitons and trions are rarely present under equilibrium conditions.
However, in narrow-gap or semimetallic systems, strong Coulomb interactions can result in exciton binding energies ($\Eex$) that are larger than the bandgap ($\Eg$) and lead to the spontaneous formation of excitons even in equilibrium. When these spontaneously-formed excitons condense, they can drive a phase transition into a novel collective state, giving rise to the excitonic insulator (EI) phase \cite{jerome_excitonic_1967}.  Dimensionality plays a crucial role in stabilizing these bound states, as it strongly influences the dielectric environment, potentially enhancing Coulomb interaction and thus enlarging the exciton binding energy \cite{deslippe_electronhole_2009,chernikov_exciton_2014,wang2018colloquium}. Several materials have been proposed as realizations of the excitonic insulator phase, though unambiguous experimental confirmation remains limited.

An alternative route to an excitonic insulator is through carefully engineered systems.  Using such devices it is possible to manipulate the exciton properties \cite{thureja_electrically_2022}, prolong their lifetimes \cite{tang_long_2019}, or use them as spectroscopic tools to probe other material properties \cite{shimazaki_strongly_2020}. Remarkably, in some engineered systems, carefully tailored environments create conditions favorable for the formation of an excitonic insulator, allowing excitons to condense spontaneously even without external excitation \cite{ma_strongly_2021,chen_excitonic_2022,wang_evidence_2019,nguyen_perfect_2025,gu_dipolar_2022}.

\begin{figure}
    \centering
    \includegraphics[width=1\linewidth]{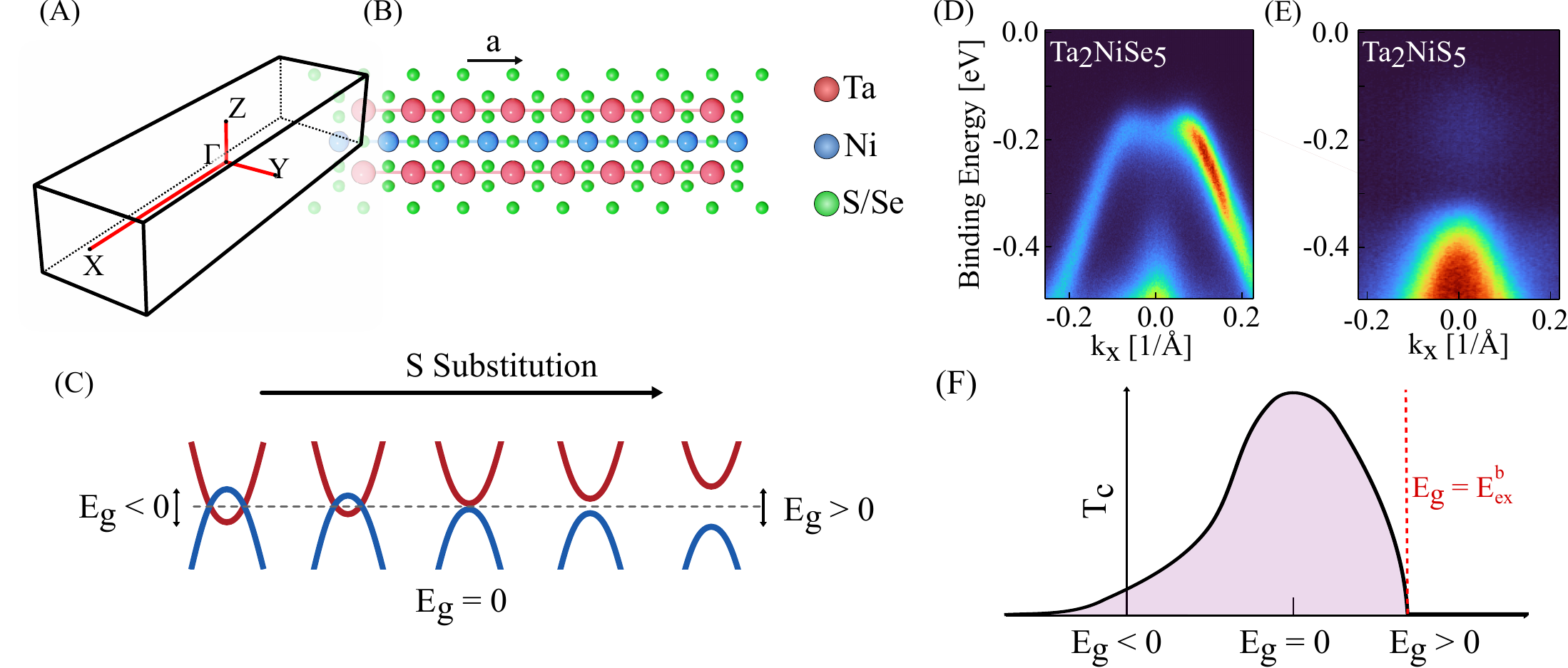}
    \caption{\textbf{Sulfur substitution in the excitonic insulator candidate $\mathrm{Ta_2Ni(Se_{1-x}S_x)_5}$:} \textbf{(A)}~Brillouin zone of $\mathrm{Ta_2Ni(Se_{1-x}S_x)_5}$, with high-symmetry points indicated. The $\Gamma$--$X$ direction corresponds to the quasi-1D chain direction shown in panel~(B). 
\textbf{(B)}~Simplified crystal structure of $\mathrm{Ta_2Ni(Se_{1-x}S_x)_5}$. Each unit cell contains one Ni atom (blue) and two Ta atoms (red), coordinated by Se/S atoms. The atoms are arranged in chains along the crystallographic $a$-axis, giving rise to a quasi-one-dimensional structure
\textbf{(C)}~Schematic illustration of the band structure evolution in $\mathrm{Ta_2Ni(Se_{1-x}S_x)_5}$ as sulfur gradually replaces selenium, highlighting the transition from a semimetallic to a semiconducting state.
\textbf{(D)}~ARPES image of the top of the valence band in \TNSE, exhibiting the characteristic ``M''-shaped dispersion commonly associated with an excitonic insulator~\cite{seki_excitonic_2014}. 
\textbf{(E)}~ARPES image of the fully sulfur-substituted compound \TNS, showing a shift of the valence band to higher binding energies. A faint in-gap feature is observed at approximately $165\,\mathrm{meV}$. 
\textbf{(F)}~Schematic phase diagram of an excitonic insulator (adapted from~\cite{jerome_excitonic_1967}). In the semiconducting regime, increasing the band gap reduces the excitonic condensation temperature until a critical point is reached at which the exciton binding energy equals the band gap \textit{$(\Eg=\Eex)$}, resulting in $T_c = 0$.}
    \label{fig:fig1}
\end{figure}

Trions having significant binding energies have been observed mainly in finely tuned two-dimensional systems, such as monolayer transition metal dichalcogenide \cite{mak2013tightly} as well as in 1D systems like carbon-nanotubes \cite{liang_solid_2016,jakubka_trion_2014,park_observation_2012}.

One of the candidate materials proposed to host an excitonic-insulator ground state is \TNSE. It is a quasi-one-dimensional system composed of Ta and Ni chains running along the crystallographic $a$-axis, resulting in a highly anisotropic Brillouin zone (see Fig.~\ref{fig:fig1}(A–B)).
There is ongoing debate over whether \TNSE\ is truly an excitonic insulator. While several studies present evidence supporting this scenario~\cite{lu_zero-gap_2017, wakisaka_excitonic_2009, wakisaka_photoemission_2012}, with evidence from angle resolved photoemission spectroscopy (ARPES) showing a phase transition leading to an M-shaped gap (see Fig.~\ref{fig:fig1}(D)), the structural transition that accompanies the electronic phase change complicates the interpretation of experimental observations~\cite{watson_band_2020}.

Upon substituting sulfur for selenium, the system evolves from a semimetal into a narrow-gap semiconductor (Fig. \ref{fig:fig1}(C)). In Fig. \ref{fig:fig1}(E and D) we show the ARPES images of the top of the valence band of \TNSE \ and \TNS, respectively. The transition temperature decreases and eventually vanishes at a critical Se/S ratio, where $\Eex=\Eg$ \cite{lu_zero-gap_2017, volkov_failed_2021, chen_anomalous_2023} (Fig. \ref{fig:fig1}(F)). \TNS, the fully substituted compound, is generally regarded as a conventional semiconductor with a direct bandgap of a few hundreds meV\cite{chen_anomalous_2023,li_strong_2017,pal_pressure-dependent_2024} presumably having $\Eex<\Eg$.
\\
However, the apparent simplicity of \TNS \ invites further scrutiny. The strong Coulomb interactions are still present and excitonic fluctuations could persist. 
In this paper we are using ARPES to study in detail the low-energy spectral features in \TNS.
We identify an unusual in-gap state that is sharply localized in momentum space, an observation that cannot be accounted for by conventional band structure theory. To explain this feature, we propose a theoretical model based on the formation of negative trions, which captures the key characteristics of the in-gap state. This interpretation is supported by direct measurements of the conduction band, confirming that the energetic conditions required for trion formation are met. Additional support comes from surface doping experiments, which reveal a systematic evolution of the in-gap spectral weight, consistent with a trion-based mechanism. Finally, detailed spectral function calculations within the trion framework provide a comprehensive description of the experimental data.

\section{Results}
\begin{figure}
    \centering
    \includegraphics[width=1\linewidth]{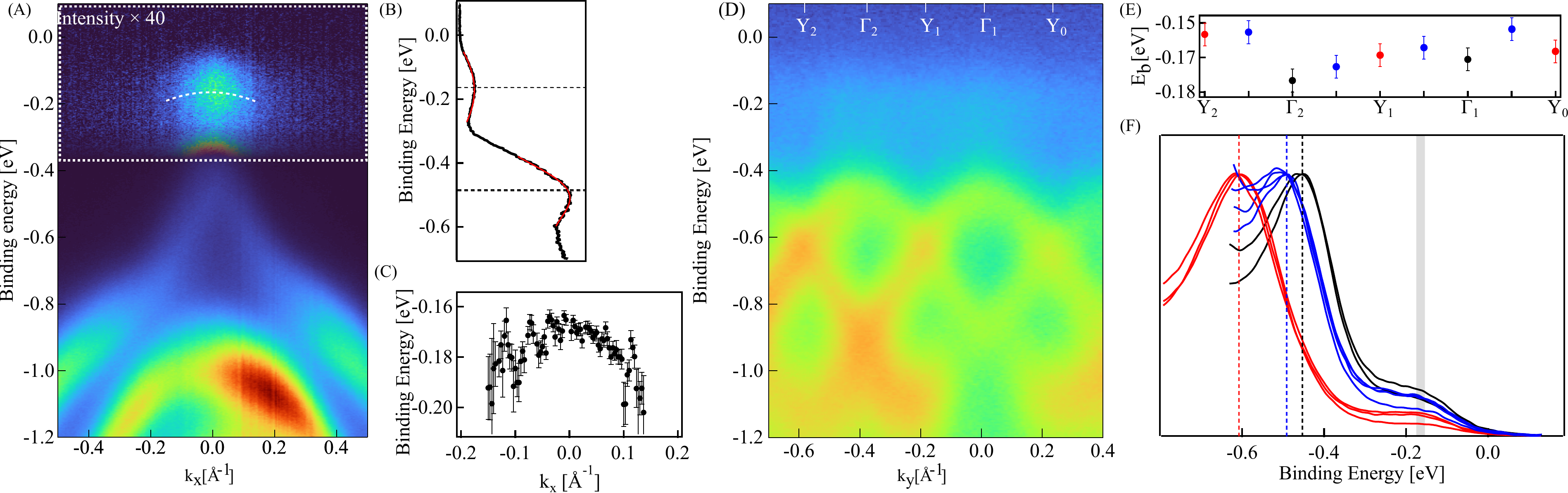}
   \caption{\textbf{Characterization of the in-gap state in Ta$_2$NiS$_5$:}
\textbf{(A)} ARPES spectrum measured with 22\,eV photons along the chain direction ($\Gamma$–X), revealing a faint and compact in-gap state located $\sim165$\,meV below the Fermi level. The intensity in the marked region (white dotted line) is enhanced by a factor of 40 for visibility. The overlaid curve shows the dispersion of the in-gap state along k$_x$, as extracted from panel (C).  
\textbf{(B)} Energy distribution curve (EDC) at the $\Gamma$ point, clearly showing the in-gap state energetically separated from both the valence band and the Fermi level. The binding energies of the in-gap state and valence band maximum are indicated by dashed lines.  
\textbf{(C)} Dispersion of the in-gap state along k$_x$, extracted by fitting the EDC peaks. Error bars represent the uncertainty in the extracted peak positions. A clear hole-like dispersion is observed, with spectral weight abruptly vanishing beyond $\pm0.2$\,\AA$^{-1}$. Data are shown only where the fits are reliable.  
\textbf{(D)} ARPES spectrum along $\Gamma$–Y (50\,eV photons, linear horizontal polarization), i.e., perpendicular to the chains. The spectrum is shown on a logarithmic scale and has been FFT-filtered for clarity. The same in-gap state is present but remains completely flat, underscoring its strong one-dimensional character.  
\textbf{(E)} Binding energy of the in-gap state measured at various momenta across multiple Brillouin zones. Energies are extracted from Gaussian fits to the EDCs; error bars indicate fitting uncertainties. The dispersion remains flat within experimental resolution.  
\textbf{(F)} EDCs corresponding to the momentum points shown in panel (E). The valence band exhibits a clear dispersion, while the in-gap state remains fixed in energy. Dashed vertical lines mark the valence band peak positions, and the shaded gray band highlights the energy range of the in-gap state.}

 \label{fig:kxyz}
\end{figure}

\subsection{In-gap state}
We begin by examining the electronic structure of \TNS.
Figure~\ref{fig:kxyz} shows ARPES spectra of \TNS, along the Ta-Ni chain direction (A)  and perpendicular to the chains (D). The top of the valence band at the $\Gamma$ point appears at a binding energy of approximately 420\,meV, consistent with previous reports~\cite{chen_anomalous_2023}. The effective mass of the uppermost hole-like band is about $0.3\pm0.04m_e$.

Upon zooming into the energy gap region, displayed with an intensity scale 40 times larger in Fig.~\ref{fig:kxyz}(A), we observe a faint, sharply localized spectral feature centered around 165\,meV below the Fermi level at the $\Gamma$ point.
In Fig.~\ref{fig:kxyz}(B) we show the energy distribution curve (EDC) at the $\Gamma$-point, the in-gap state remains clearly separated from the valence band, yet does not intersect the Fermi level, indicating its unusual character. 

 In Fig.~\ref{fig:kxyz}(C), we present the dispersion of the in-gap state along the chain direction, extracted by fitting Gaussian functions to the EDCs. We observe a hole-like dispersion confined within $\pm 0.2\,\text{\AA}^{-1}$ along the one-dimensional chain direction, beyond which the spectral intensity drops off sharply (see SM for details of the fitting procedure).
The bandwidth of the in-gap state is approximately 20\,meV, and a parabolic fit yields an effective mass of about $3\,m_e$, roughly an order of magnitude larger than the effective mass of the valence band.

 In Fig. \ref{fig:kxyz}(D) we show the dispersion perpendicular to the chains direction.  The effective mass of the valence band in this direction is about 3 times larger than along the chains. The in-gap state appears nearly dispersionless and delocalized in momentum space.
The binding energy of the in-gap state, presented in Fig. \ref{fig:kxyz}(E) at the $\Gamma$ and Y points (and in between), across multiple Brillouin zones, remains constant, confirming the absence of dispersion perpendicular to the chains. Binding energies were extracted from EDC's shown in Fig.~\ref{fig:kxyz} (F).
Similar behavior is observed for the dispersion out-of-plane ($\Gamma$-Z) (see SM).

We noticed that the spectral weight of this anomalous in-gap state varies significantly between cleaves, sometimes prominent and well-defined, other times nearly invisible. A weak in-gap feature was in fact reported previously by Chiba \textit{et al.}~\cite{chiba_valence-bond_2019}, who attributed a small spectral weight around $-0.2$~eV to a surface state within the band gap. In contrast, in other ARPES studies this feature is not resolved~\cite{chen_anomalous_2023}, likely due to its low relative intensity and strong sensitivity to surface conditions.
However, we emphasize that whenever the in-gap state is observed, its binding energy and momentum-space characteristics remain remarkably consistent, supporting its intrinsic origin.

We find that the in-gap state shows only a weak temperature dependence between 20K and room temperature (see SM for more details).

\subsection{Energetics and trion interpretation}
The central question that emerges is: what is the microscopic origin of this unusual in-gap state?
One possible interpretation is that the in-gap state originates from excitons, particularly given Ta$_2$NiS$_5$'s proximity to the excitonic-insulator phase observed in the compound Ta$_2$Ni(S$_x$Se$_{x-1}$)$_5$.
While excitons cannot be directly probed by photoemission, ARPES can nonetheless provide insights into their presence and properties through their influence on the single-particle spectral function.
In particular, theoretical frameworks have been developed to interpret excitonic signatures in time-resolved ARPES spectra \cite{rustagi_photoemission_2018}. A photoemission process that dissociates an exciton should produce a spectral feature located \( \Eex \) below the bottom of the conduction band, since this is the additional binding energy that must be supplied to break the exciton apart.  Signatures of excitons have been provided by pump-probe ARPES experiments \cite{Dani_dark_exciton,mori_spin-polarized_2023}. 
Mahan excitons that can be created in metals \cite{mahan_excitons_1967} were also measured using ARPES \cite{ma_multiple_2022}. 
The characteristic "ball"-like spectral shape of our in-gap state closely resembles the excitonic fingerprint observed in time-resolved ARPES experiments in TMDs \cite{Dani_dark_exciton,Iliya_review} and Topological Insulators \cite{mori_spin-polarized_2023}.
Of particular relevance for this work is the claim of a signature of excitons at temperatures above the phase transition in \TNSE  \cite{fukutani_detecting_2021}.

However, a key distinction must be noted. While \TNSE\ is widely regarded as an excitonic insulator, \TNS\ is not. Therefore, in \TNSE\  the exciton binding energy, $\Eex$, is expected to exceed the band gap, $\Eg$, implying that a photoemission process that dissociates an exciton should produce a spectral feature located {\em below} the valence band maximum. Such a feature has indeed been observed in ARPES experiments \cite{fukutani_detecting_2021}, although its interpretation as a signature of excitons remains a subject of ongoing debate \cite{Yale_first}.
In the case of \TNS\ studied here, attributing the 165 meV in-gap feature to an exciton would result in $\Eg>\Eex$, in agreement with it being a semiconductor, which should make thermally excited excitons very rare. The very weak temperature dependence of the in-gap state indicates that it is not governed by a thermally excited exciton population.

We propose that the in-gap feature observed in our measurements originates from trions: three-body bound states consisting of an exciton bound to an additional electron. These trions form in the presence of excess surface charge, offering a natural explanation for the observed spectral signature as we discuss below. Unlike simple excitonic models, which are incompatible with the energetics of the system, the trion scenario resolves this contradiction by invoking a more strongly bound composite state, stabilized by the extra electron–exciton interaction, \( \Eexe \). The total trion binding energy, \( \Etr =\Eex + \Eexe \), can then exceed the band gap even if the exciton binding energy \( \Eex \) alone does not. In such cases, individual excitons cannot form spontaneously, but the presence of an additional surface electron enables the formation of a trion. 

\begin{figure}
    \centering
    
    \includegraphics[width=1\textwidth]{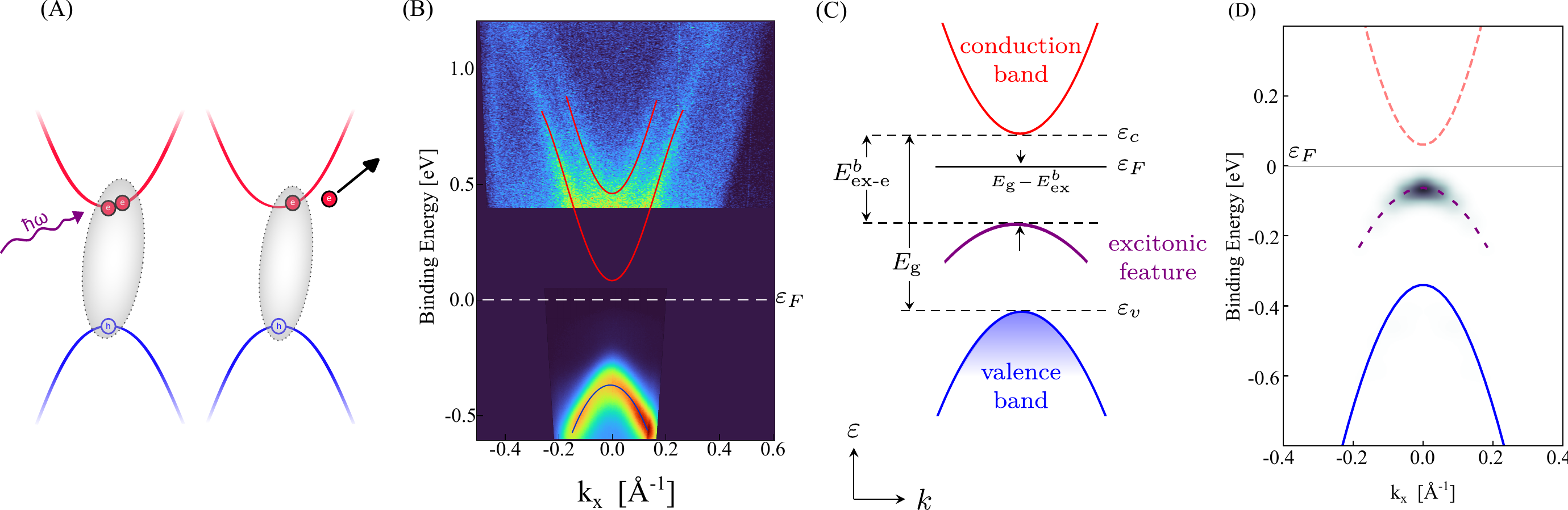}
    \caption{\textbf{Photoemission process and energetics:} \textbf{(A)} Schematic of the photoemission process involving a trion. Left: Initial ground state where an exciton and a conduction-electron form a trion bound state. Right: upon photon absorption, one electron is emitted, breaking the trion and leaving behind a neutral exciton. 
    \textbf{(B)} The 6.4eV ARPES spectra showing the top of the valence band, combined with the 6eV 2PPE spectra showing the bottom of the conduction band. The solid lines are fits to the data using a simple model. More details about the model can be found in the SM.  
    \textbf{(C) } Illustration of the band structure and the the in-gap feature resulting from the exciton remaining after breaking a trion. Conduction band (red), valence band (blue) and in-gap feature (magenta). The excitonic binding energy ($\Eex$) and the electron-exciton binding energy ($\Eexe$) are indicated. 
    \textbf{(D)} Calculated spectral function obtained in the presence of a single trion, due to photoemission of a conduction electron, obtained using exact diagonalization (ED) on the minimal model presented in the text. To mimic the finite temperature and resolution of the ARPES measurement we use convolution with a Gaussian of width $20$meV.}
    
    \label{fig:Energetics}
\end{figure}

In a photoemission process, such a trion can be broken, leaving behind an exciton. This process is illustrated in Fig. \ref{fig:Energetics}(A). Similarly to the signature left by the disassociation of an exciton, we show below that the disassociation of a trion should produce a spectral feature located \( \Eexe \) below the bottom of the conduction band, since this is the additional binding energy needed to break the trion into the photo-emitted electron and the leftover exciton. 

To determine the exciton and trion binding energies from the ARPES spectra, it is therefore necessary to first estimate the single-particle gap: the energy difference between the bottom of the conduction band and the top of the valence band, illustrated in \ref{fig:fig1}(C). 
For this purpose, we employed two-photon photoemission (2PPE) to probe the unoccupied electronic states. In Fig. \ref{fig:Energetics}(B) we show the 2PPE data combined with ARPES data, showing the single particle gap. The measurement reveals two parabolic bands. By applying a simple fitting model, we estimate that the bottom of the conduction band, $\Ec$, lies approximately 85meV $\pm$ 20meV above the Fermi level. More details are provided in the SM.
The top of the valence band is found at a binding energy of 420 meV$\pm$40 meV. Yielding a band gap of about 500 meV $\pm$ 40 meV, comparable to previous reports \cite{chen_anomalous_2023,mu2018electronic}. 

We now turn to a more careful analysis of the energetics. In the experimental setup, the system is coupled to a charge reservoir that fixes the chemical potential. In this situation, an electron can be transferred from the reservoir into the conduction band. In a non-interacting system, this occurs when the bottom of the conduction band, $\Ec$, crosses the Fermi level, $\Ef$. However, interactions can significantly modify this condition.

In particular, the formation of a bound trion state can render charge transfer energetically favorable even when $\Ec > \Ef$. The key requirement is that the total binding energy of the trion, $\Etr = \Eex + \Eexe$, exceeds the sum of the single-particle gap $\Eg$ and the energy cost $(\Ec - \Ef)$ of placing an electron in the conduction band. That is, charge transfer can form when:
\begin{equation}
\Etr \geq \Eg + (\Ec - \Ef)
\label{E_C}
\end{equation}
Equivalently, Eq.~\eqref{E_C} can be viewed as a condition on the Fermi level $\Ef\geq\Ec+\Eg-\Etr$ for the charge transfer to occur. 
We note that, although the Fermi level is defined as the energy required to add a single electron to the system, in this case its addition also induces the formation of an exciton, which subsequently binds to the added electron.

 In the presence of a \emph{single trion} (or more realisticaly - a low density of trions), a photoemission process that breaks it will result in a spectral feature at energy $E^b_\text{ex-e}$ below the bottom of the conduction band or, equivalently, $\Eg - \Eex$ below the Fermi level. This energy balance is illustrated in Fig.~\ref{fig:Energetics}(C).

Based on this analysis and the positions of the valence and conduction bands, as well as the position of the in-gap feature, observed experimentally we can extract the binding energies in the system, yielding $\Eex = 340\pm60$ meV and $\Eexe=250\pm50$  meV .

\subsection{Theoretical model and spectral function calculation}

To analyze the formation of trions in the system and the expected spectral function in their presence in a more quantitative manner, we study a minimal 1D lattice model for the weakly-coupled \TNS\ chains accounting for the geometry of the unit cell.

A schematic picture of the model is shown in the SM in Fig.~\ref{fig:model}(A). We consider two chains of conduction electrons, labeled by $\alpha=1,2$, and a single chain of valence electrons, corresponding to the two Ta and the single Ni chains respectively. For simplicity we consider a spinless model. The Hamiltonian is given by
\begin{align}\label{eq:model}
    H &= H_0+H_{\rm int} \nonumber \\
    H_0
    &=
    -t_c \sum_{\alpha,i} c^\dagger_{\alpha,i} c_{\alpha,i+1}
    +t_f \sum_{i}f^\dagger_{i} f_{i+1}
    +D \sum_{i} \Big( n^c_{i} - n^f_{i}\Big)
    -\mu \sum_{i} \Big( n^c_{i} + n^f_{i}\Big) \nonumber
    \\
    H_\text{int}
    &=  \sum_{\alpha, i,j} U_{ij}
    n^c_{\alpha,i} n^f_{j}
    + \sum_{\alpha,\beta,i,j} V^{\alpha\beta}_{ij}
    n^c_{\alpha,i} n^c_{\beta,j} .
\end{align}
Here $c_{\alpha,i}$ is the annihilation operator of a conduction electron on site $i$ on chain $\alpha=1,2$, $f_{i}$ is the annihilation operator of a valence electron at site $i$, $t_c$ ($t_f$) is the intra-chain hopping amplitude on the conduction (valence) chain, $D$ determines the single particle gap $E_\text{g}=D-2(t_f+t_c)$, $\mu$ is the chemical potential, $U_{ij}$ is the interaction between the conduction and valence electrons, $V^{\alpha\beta}_{ij}$ is the interaction between the conduction electrons on chains $\alpha,\beta$, $n^c_{\alpha,i}$ ($n^f_i$) is the respective $c_\alpha$ ($f$)-electron number operator, with $n^c_i=\sum_\alpha n^c_{\alpha,i}$ the total density of conduction electrons.   
Model parameters used in the simulations are $t_c=0.78$eV and $t_f=1.09$eV, for the hopping amplitudes, corresponding to the effective masses of the conduction and valence bands, respectively. $c$ and $f$ hybridization is forbidden in orthorhombic phase \cite{watson_band_2020}.
Hopping between the two conduction chains is neglected, consistent with previous studies showing Ta-Ta chain coupling to be below $0.02$eV~\cite{mazza_nature_2020,kaneko_orthorhombic--monoclinic_2013,yamada_fflo_2016,sugimoto_strong_2018,PhysRevB.94.085111}.
In addition, we set $D=4.14$eV resulting in a single particle gap equal to $\Eg=0.4$eV.

The interactions $U_{ij}$ and $V^{\alpha\beta}_{ij}$ are modeled using a screened Coulomb potential and their explicit form is given in the SM.
While interactions between the conduction ($c$) and valence ($f$) electrons enhance both exciton and trion binding energies, interactions between conduction electrons suppress the latter. Trion formation is thus energetically favorable only if the attraction of the extra electron to the exciton’s hole outweighs the repulsion from its electron.
Note also that we do not include interactions between the valence electrons in the model as we will only consider states within the single-hole occupancy subspace.

To obtain the exciton and trion binding energies we study the model using exact diagonalization (see SM for details). 
We find that the exciton-electron binding energy, $\Eexe$, in the system can be significant in comparison to the exciton binding energy, with their ratio reaching a value close to one third as observed in the experiment. 
Specifically, for screening length $\xi=6 a$ (with $a=3.41$\AA\ the unit cell length~\cite{jain_commentary_2013}), as used in the calculation of the spectral function discussed below, we obtain $\Eex\approx 0.34$eV and $\Eexe\approx0.12$eV.

We next calculate the spectral function probed by the photoemission process. Focusing on the effect of the presence of trions in the system, and assuming zero temperature, we calculate 
\begin{equation}
    A_{c}(k,\omega)=\sum_{\alpha} \sum_n \big|\langle \psi_n|c_{\alpha,k}|\psi_\text{gs}\rangle\big|^2\delta\big(\omega-(E_n-E_\text{gs}) \big),
\end{equation} 
where $|\psi_\text{gs}\rangle$ is the many-body ground state wavefunction, $E_\text{gs}$ is the ground state energy, $|\psi_n\rangle$ is the wavefunction in an excited state, and $E_n$ is the respective energy. 
Here, we  only consider processes in which a conduction electron is emitted from the system. We expect photoemission of valence electrons to predominantly probe the valence band and therefore forgo an explicit calculation of this process.

Assuming the density of trions is low, and to elucidate the analysis, we consider the scenario of a single trion in the ground state.
A photoemission process of a conduction electron requires breaking the trion leaving behind an electron-hole pair in the system. 
Thus to obtain the spectral function we carry out a summation over excited states $|\psi_n\rangle$ in the subspace with a single electron and a single hole. We find that the largest matrix element is with states hosting a bound exciton at momenta close to zero, resulting in the ball-like feature as shown in Fig.~\ref{fig:Energetics}(D).
Strikingly, the resulting spectrum closely matches the in-gap feature observed in our ARPES data.

\subsection{Surface doping as a control knob}
The density of trions is expected to depend on the level of surface doping.
To investigate this dependence, we examined the evolution of the in-gap state under two conditions:
(1) by allowing the sample to age in ultra-high vacuum (UHV), and
(2) by depositing potassium on the surface to introduce controlled electron doping.

\begin{figure}
    \centering
    \includegraphics[width=0.75\linewidth]{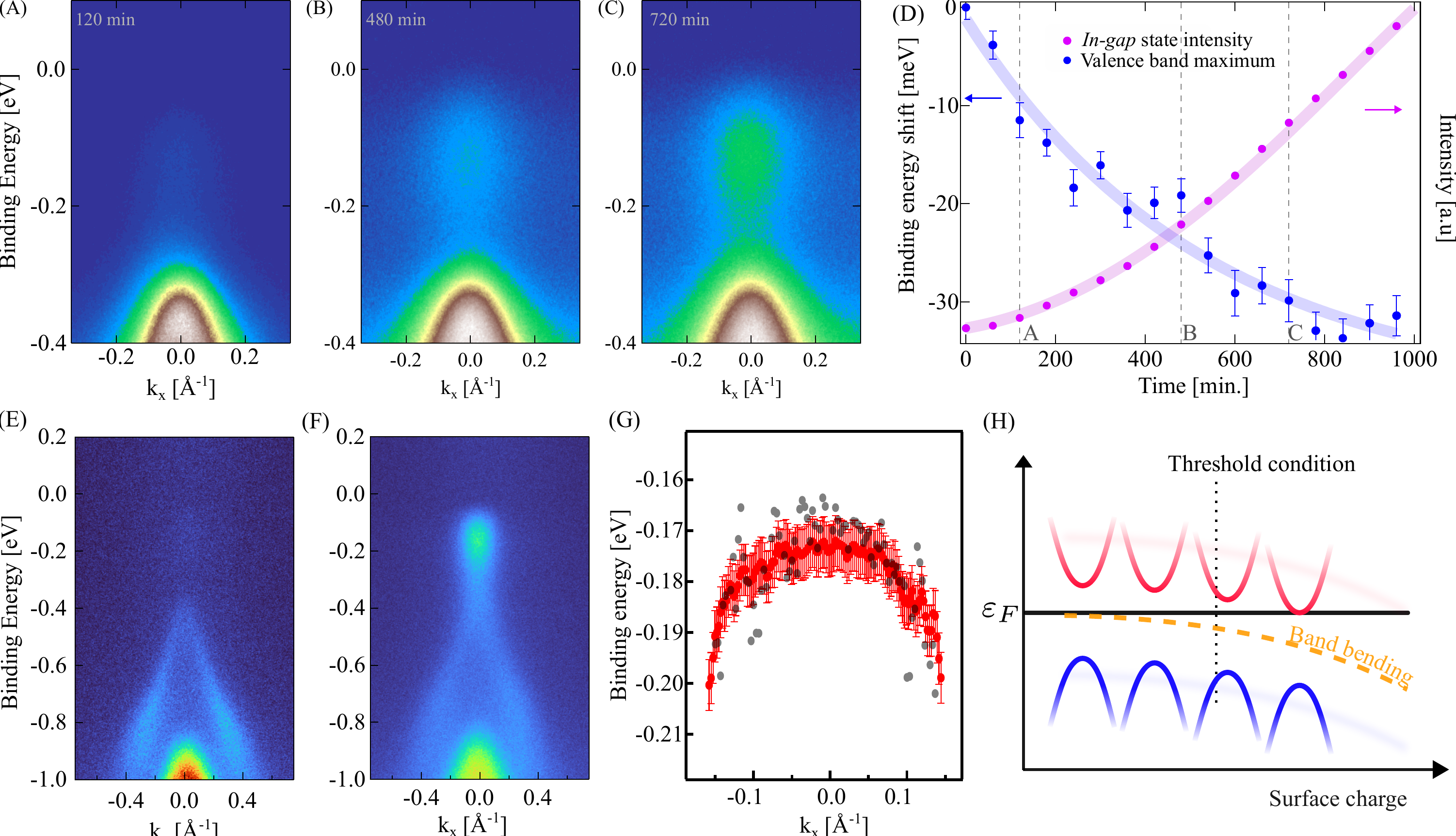}
    \caption{
    \textbf{Surface doping dependence of the in-gap state:} 
    \textbf{(A–C)} ARPES spectra along the $\Gamma$–X direction measured 
at successive time intervals after cleaving the sample using 6.4\,eV photons. 
    (A) Two hours, no in-gap state is visible.
    (B) After 8 hours in vacuum, a faint in-gap feature begins to emerge. 
    (C) After 12 hours, the spectral weight of the in-gap state increases significantly. 
    \textbf{(D)} Time evolution of the valence band maximum (blue) and the spectral intensity of the \textit{in-gap} state(magenta). Shaded lines serve as guides to the eye. EDCs at the $\Gamma$ point as a function of time used here are shown in the SM. Dashed vertical lines indicate the time points at which panels A–C were measured. Intensity values are normalized to the intensity of the peak of the valence band.
    \textbf{(E–G)} Controlled potassium surface doping of Ta$_2$NiS$_5$. 
    (E–F) ARPES spectra along the $\Gamma$–X direction, taken with 43\,eV photons and horizontal polarization, measured before (E) and after 20 seconds of potassium deposition (F). 
    (G) Extracted dispersion of the in-gap state from panel (F) (red), overlaid with the dispersion from Fig.~\ref{fig:kxyz}(C) (black) for comparison. 
    \textbf{(H)} Schematic illustration of the effect of surface doping: band structure shifts downward, and once  the condition in Eq. \ref{E_C} is satisfied, trion formation becomes energetically favorable.}

    \label{fig:surface_doping_combined}
\end{figure}

 Fig.~\ref{fig:surface_doping_combined} (A-C) show ARPES spectra along the chains direction, measured with 6.4~eV photons at successive time intervals after cleaving the sample. At early times (panel A), the in-gap state is barely visible. As time progresses (panels B and C), its intensity steadily increases. Simultaneously, the valence band shifts toward higher binding energies, indicating a change in the surface potential. Previous studies have shown that such binding energy drifts over time are related to changes in surface potential \cite{king_large_2011, bianchi_electronic_2012}. Panel D quantifies this evolution: the energy shift of the valence band maximum (blue) and the integrated intensity of the in-gap state (magenta) are plotted as a function of time. The growing in-gap intensity, reflecting an increasing surface trion density, tracks the downward band bending. This correlation supports our interpretation that the in-gap spectral weight is directly linked to the level of surface doping, consistent with the trion-based model.

A more controlled method for electron doping of the surface is in-situ potassium deposition on the freshly cleaved sample. In Fig.~\ref{fig:surface_doping_combined}(E), we show the ARPES spectrum acquired immediately after cleaving, where the spectral weight within the gap is very weak.
Panel (F) displays the spectrum following the deposition of approximately one-tenth of a monolayer of potassium. After deposition, we observe a slight downward shift of the valence band and a pronounced enhancement of the in-gap state intensity. Notably, the dispersion of the in-gap state post-deposition, shown in Fig.~\ref{fig:surface_doping_combined}(G), is identical to that observed in aged samples, indicating a common origin.
The effect of surface doping is illustrated schematically in Fig.~\ref{fig:surface_doping_combined}(H). Doping causes a rigid downward shift of the bands. At a certain doping level, the conduction band approaches the Fermi level closely enough that Eq.~\ref{E_C} is satisfied, leading to the formation of the first trion.

At lower photon energies, the increased probing depth averages over surface and bulk regions with different band bending, reducing the spectral weight of the in-gap state and broadening its lineshape. This behavior is consistent with trion formation being restricted to the surface region where band bending stabilizes them.

\newpage
\clearpage

\section{Discussion}
Let us now consider alternative origins to the in-gap state. Can the in-gap state represent the bottom of the conduction band? We argue that this interpretation is not applicable to \TNS, for two reasons. First, although DFT calculations for this system are challenging due to strong Coulomb interactions, there is consensus that \TNS\ is \emph{not} semimetallic \cite{Rubio_DFT}, implying that the conduction band minimum lies above the Fermi level. This is in agreement with our 2PPE measurements, which directly reveal the unoccupied states. Moreover, the observed in-gap state exhibits a relatively heavy, hole-like dispersion and is never seen to cross the Fermi level, inconsistent with a conduction band origin.

Additionally, we argue that the in-gap state does not originate from an impurity band. Observing an impurity band in ARPES would require a high concentration of defects, for which we find no evidence. On the contrary, Raman spectroscopy reveals sharp phonon modes, indicating low disorder \cite{ye2021lattice}. Moreover, we observe consistent spectra across samples from different growth batches, further supporting sample uniformity. Finally, the systematic evolution of the in-gap state intensity with potassium doping is inconsistent with the behavior expected from an impurity band.

How do the numbers extracted from the ARPES data using the trion model compare with existing literature?  ARPES data at high temperature reveal a band gap of $415\,\mathrm{meV} \pm 20\,\mathrm{meV}$~\cite{chen_anomalous_2023}. Similarly, high-temperature transport measurements report a gap of approximately 500meV~\cite{lu_zero-gap_2017}, in agreement  with our estimate.
 Excitonic binding energies on the order of hundreds of meV have been observed in various low-dimensional materials. For example, binding energies of approximately 240\,meV have been reported in monolayers of MoS$_2$ and WSe$_2$~\cite{park2018direct}, and values as high as 700\,meV have been observed in monolayer WS$_2$~\cite{zhu2015exciton}. In carbon nanotubes, binding energies exceeding 400\,meV have been measured~\cite{maultzsch2005exciton}, consistent with theoretical predictions that strong one-dimensional quantum confinement leads to significantly enhanced exciton binding energies~\cite{brown_exciton_1987}.
Most notably, optical measurements of \TNS\ indicate an exciton binding energy of approximately  $\alpha 
\cdot160meV$  where $\alpha$ is of order unity\cite{larkin2017giant} in agreement with our result.

The trion binding energy, $E^b_{ex-e}$,calculated from our ARPES data to be approximately 250 meV, is relatively large but consistent with expectations for quasi-1D systems \cite{brown_exciton_1987} with reduced screening\cite{chernikov_exciton_2014}.
Trion binding energies of up to 200meV were found in carbon nanotubes \cite{mohl2018trion}.

The quasi-one-dimensional (quasi-1D) structure of \TNS\ provides a natural setting for strong electronic correlations.
The localization of the trion state in the out-of-chain directions, evidenced by its flat dispersion along k$_y$ and k$_z$, suggests that the trion is confined primarily to individual chains. Assuming a lateral confinement width of approximately $d \sim 4$\,\AA, corresponding to the structural width of a single unit-cell in \TNS\cite{jain_commentary_2013}, the observed trion binding energy is consistent with theoretical predictions for trions in quasi-1D systems\cite{park_observation_2012}.

The presence of the double Ta chain, which forms the conduction band, further enhances the exciton-to-trion binding energy ratio. The negligible Ta–Ta coupling allows the two electrons in a trion to localize on different chains, increasing their average separation relative to the electron–hole distance. This reduces the Coulomb repulsion within the trion and enhances its binding energy relative to that of the exciton (see SM).


Trion-related features in ARPES have been reported in TaSe$_3$ \cite{ma_multiple_2022} and in a doped monolayer of WS$_2$ \cite{Exciton_ARPES_WS2}. However, the case of \TNS \ is fundamentally different in two key respects. In those earlier systems, the bound states form in the presence of a Fermi surface, which provides a supply of low-energy carriers, essentially "cheap" particle-hole pairs, that can readily bind to the photo-hole created during the photoemission process. These are not true equilibrium trions, but transient final-state effects that manifest in the spectral function.
In contrast, we argue that the in-gap state observed in \TNS \ reflects a genuine equilibrium trion population. First, the state is clearly visible even in the absence of a Fermi surface, ruling out the possibility of photo-hole binding to ambient carriers. Second, \TNS \ is a direct-gap semiconductor, which makes the formation of long-lived "dark" excitons, generated by one photon and detected by another, unlikely. Together, these considerations support an interpretation in which the observed in-gap feature originates from pre-existing trions in thermal equilibrium.

\section{Summary}
In summary, we report the discovery of a distinct \textit{in-gap} state in \TNS, whose properties are incompatible with those of conventional electronic or excitonic bands. Based on a comprehensive analysis combining ARPES measurements, controlled surface doping, and theoretical modeling, we identify this feature as a trion, a three-body bound state consisting of an exciton and an additional electron. Unlike excitons, trions can remain energetically stable even when the exciton binding energy \( E_{\mathrm{ex}}^b \) is smaller than the band gap \( E_g \), provided that the binding energy between the exciton and the additional electron is sufficiently large. Surface doping plays a crucial role by supplying the excess electrons necessary to enable trion formation. Our results demonstrate that interaction-driven surface quasiparticles can emerge in otherwise conventional semiconductors, and establish \TNS\ as a rare platform where spontaneous trion formation can be both stabilized and experimentally tuned.

\newpage

\section*{Methods}
High quality Ta$_2$NiSe$_{5-x}$S$_x$ single crystals were synthesized  via  
chemical vapor transport (CVT) method. 
The crystallinity of samples was confirmed  by XRD measurements and the elemental composition was determined through energy dispersive X-ray(EDX) analysis.\\

Photoemission experiments were performed at the ULTRA beamline at SLS (Paul Scherrer Institute, Viligen, Swizterland) and the U112-PGM and U125-PGM beamlines at BESSY-II (ARPES 1$^2$ and spin-ARPES setups, HZB, Berlin, Germany), and the Cassiopee beamline at Soleil (Paris, France).
Samples were cleaved in-situ at base temperature of about 25K. Given the insulating nature of the sample, particular care was taken to suppress charging, achieved by operating at low photon flux.

Potassium was deposited onto the sample surface using a calibrated getter source. Based on the getter calibration, 20sec deposition time amounts to a 10\% of a monolayer coverage. 

Low photon energy data were measured in the lab at the Technion using a 6.4 eV linear horizontal polarized laser with a pulse duration of 2ps.

Two-photon photoemission (2PPE) measurements of the conductance band were performed in the lab at TAU with ultrashort (~40fs) 6eV laser pulses and linear horizontal polarization \cite{SOBOTA}. To increase the visible momentum range on the detector, a bias of -30V was applied between the sample stage and entrance cone of the hemispherical analyzer (DA30L, Scienta) \cite{Bias}. The sample was kept at 80K and at a pressure better than 1.5 10$^{-10}$ mbar throughout the measurement. A 2PPE measurement requires a much higher intensity beam than used in equilibrium ARPES \cite{SOBOTA}. Such intensity would result in an overwhelming signal from occupied bands, and the low energy cutoff of the presented spectrum was set to avoid excessive counts.

\section*{Data Availability} 
The data that support the findings of this study are available from the corresponding
authors author upon reasonable request.


\section*{Acknowledgments}
We acknowledge the Paul Scherrer Institute, Villigen, Switzerland for provision of synchrotron radiation beamtime at beamline SIS of the SLS.  We thank the Helmholtz Zentrum Berlin for the allocation of synchrotron radiation beamtime. We acknowledge SOLEIL for provision of synchrotron radiation facilities.
A.K.\ acknowledges funding by the Israel Science Foundation (Grant No.\ 2443/22).

\bibliography{Quantum_ball}

@article{bianchi_electronic_2012,
	title = {The electronic structure of clean and adsorbate-covered {Bi$_2$Se$_3$}: an angle-resolved photoemission study},
	volume = {27},
	number = {12},
	urldate = {2025-04-22},
	journal = {Semiconductor Science and Technology},
	author = {Bianchi, Marco and Hatch, Richard C and Guan, Dandan and Planke, Tilo and Mi, Jianli and Iversen, Bo Brummerstedt and Hofmann, Philip},
	month = nov,
	year = {2012},
	pages = {124001},
}

@article{king_large_2011,
	title = {Large {Tunable} {Rashba} {Spin} {Splitting} of a {Two}-{Dimensional} {Electron} {Gas} in {Bi$_2$Se$_3$}},
	volume = {107},
	doi = {10.1103/PhysRevLett.107.096802},
	number = {9},
	urldate = {2025-04-22},
	journal = {Physical Review Letters},
	author = {King, P. D. C. and Hatch, R. C. and Bianchi, M. and Ovsyannikov, R. and Lupulescu, C. and Landolt, G. and Slomski, B. and Dil, J. H. and Guan, D. and Mi, J. L. and Rienks, E. D. L. and Fink, J. and Lindblad, A. and Svensson, S. and Bao, S. and Balakrishnan, G. and Iversen, B. B. and Osterwalder, J. and Eberhardt, W. and Baumberger, F. and Hofmann, Ph.},
	month = aug,
	year = {2011},
	pages = {096802},
}

@article{park_observation_2012,
	title = {Observation of {Negative} and {Positive} {Trions} in the {Electrochemically} {Carrier}-{Doped} {Single}-{Walled} {Carbon} {Nanotubes}},
	volume = {134},
	issn = {0002-7863},
	url = {https://doi.org/10.1021/ja304282j},
	doi = {10.1021/ja304282j},
	number = {35},
	urldate = {2025-04-23},
	journal = {Journal of the American Chemical Society},
	author = {Park, Jin Sung and Hirana, Yasuhiko and Mouri, Shinichiro and Miyauchi, Yuhei and Nakashima, Naotoshi and Matsuda, Kazunari},
	month = sep,
	year = {2012},
	note = {Publisher: American Chemical Society},
	pages = {14461--14466},
}

@article{jakubka_trion_2014,
	title = {Trion {Electroluminescence} from {Semiconducting} {Carbon} {Nanotubes}},
	volume = {8},
	issn = {1936-0851},
	url = {https://doi.org/10.1021/nn503046y},
	doi = {10.1021/nn503046y},
	number = {8},
	urldate = {2025-04-23},
	journal = {ACS Nano},
	author = {Jakubka, Florian and Grimm, Stefan B. and Zakharko, Yuriy and Gannott, Florentina and Zaumseil, Jana},
	month = aug,
	year = {2014},
	note = {Publisher: American Chemical Society},
	pages = {8477--8486},
}

@article{lu_zero-gap_2017,
	title = {Zero-gap semiconductor to excitonic insulator transition in {Ta$_2$NiSe$_5$}},
	volume = {8},
	copyright = {2017 The Author(s)},
	issn = {2041-1723},
	url = {https://www.nature.com/articles/ncomms14408},
	doi = {10.1038/ncomms14408},
	number = {1},
	urldate = {2025-04-23},
	journal = {Nature Communications},
	author = {Lu, Y. F. and Kono, H. and Larkin, T. I. and Rost, A. W. and Takayama, T. and Boris, A. V. and Keimer, B. and Takagi, H.},
	month = feb,
	year = {2017},
	note = {Publisher: Nature Publishing Group},
	pages = {14408},
}

@article{wakisaka_excitonic_2009,
	title = {Excitonic {Insulator} {State} in {Ta$_2$NiSe$_5$} {Probed} by {Photoemission} {Spectroscopy}},
	volume = {103},
	url = {https://link.aps.org/doi/10.1103/PhysRevLett.103.026402},
	doi = {10.1103/PhysRevLett.103.026402},
	number = {2},
	urldate = {2025-05-07},
	journal = {Physical Review Letters},
	author = {Wakisaka, Y. and Sudayama, T. and Takubo, K. and Mizokawa, T. and Arita, M. and Namatame, H. and Taniguchi, M. and Katayama, N. and Nohara, M. and Takagi, H.},
	month = jul,
	year = {2009},
	note = {Publisher: American Physical Society},
	pages = {026402},
}

@article{mori_spin-polarized_2023,
	title = {Spin-polarized spatially indirect excitons in a topological insulator},
	volume = {614},
	copyright = {2023 The Author(s), under exclusive licence to Springer Nature Limited},
	issn = {1476-4687},
	url = {https://www.nature.com/articles/s41586-022-05567-3},
	doi = {10.1038/s41586-022-05567-3},
	number = {7947},
	urldate = {2023-03-19},
	journal = {Nature},
	author = {Mori, Ryo and Ciocys, Samuel and Takasan, Kazuaki and Ai, Ping and Currier, Kayla and Morimoto, Takahiro and Moore, Joel E. and Lanzara, Alessandra},
	month = feb,
	year = {2023},
	pages = {249--255},
}

@article{fukutani_detecting_2021,
	title = {Detecting photoelectrons from spontaneously formed excitons},
	volume = {17},
	copyright = {2021 The Author(s), under exclusive licence to Springer Nature Limited},
	issn = {1745-2481},
	url = {https://www.nature.com/articles/s41567-021-01289-x},
	doi = {10.1038/s41567-021-01289-x},
	number = {9},
	urldate = {2023-03-19},
	journal = {Nature Physics},
	author = {Fukutani, Keisuke and Stania, Roland and Il Kwon, Chang and Kim, Jun Sung and Kong, Ki Jeong and Kim, Jaeyoung and Yeom, Han Woong},
	month = sep,
	year = {2021},
	pages = {1024--1030},
}

@article{mazza_nature_2020,
	title = {Nature of {Symmetry} {Breaking} at the {Excitonic} {Insulator} {Transition}: {Ta$_2$NiSe$5$}},
	volume = {124},
	number = {19},
	urldate = {2023-03-19},
	journal = {Physical Review Letters},
	author = {Mazza, Giacomo and Rösner, Malte and Windgätter, Lukas and Latini, Simone and Hübener, Hannes and Millis, Andrew J. and Rubio, Angel and Georges, Antoine},
	month = may,
	year = {2020},
	note = {Publisher: American Physical Society},
	pages = {197601},
}

@article{watson_band_2020,
	title = {Band hybridization at the semimetal-semiconductor transition of {Ta$_2$NiSe$_5$} enabled by mirror-symmetry breaking},
	volume = {2},
	url = {https://link.aps.org/doi/10.1103/PhysRevResearch.2.013236},
	doi = {10.1103/PhysRevResearch.2.013236},
	number = {1},
	urldate = {2023-03-19},
	journal = {Physical Review Research},
	author = {Watson, Matthew D. and Markovic, Igor and Morales, Edgar Abarca and Le Fèvre, Patrick and Merz, Michael and Haghighirad, Amir A. and King, Philip D. C.},
	month = mar,
	year = {2020},
	pages = {013236},
}

@article{rustagi_photoemission_2018,
	title = {Photoemission signature of excitons},
	volume = {97},
	url = {https://link.aps.org/doi/10.1103/PhysRevB.97.235310},
	doi = {10.1103/PhysRevB.97.235310},
	number = {23},
	urldate = {2023-03-19},
	journal = {Physical Review B},
	author = {Rustagi, Avinash and Kemper, Alexander F.},
	month = jun,
	year = {2018},
	note = {Publisher: American Physical Society},
	pages = {235310},
}

@article{jerome_excitonic_1967,
	title = {Excitonic {Insulator}},
	volume = {158},
	number = {2},
	urldate = {2025-05-07},
	journal = {Physical Review},
	author = {Jérome, D. and Rice, T. M. and Kohn, W.},
	month = jun,
	year = {1967},
	pages = {462--475},
}

@article{nguyen_perfect_2025,
	title = {Perfect {Coulomb} drag in a dipolar excitonic insulator},
	volume = {388},
	number = {6744},
	journal = {Science},
	author = {Nguyen, Phuong X. and Ma, Liguo and Chaturvedi, Raghav and Watanabe, Kenji and Taniguchi, Takashi and Shan, Jie and Mak, Kin Fai},
	month = apr,
	year = {2025},
	pages = {274--278},
}

@article{gu_dipolar_2022,
	title = {Dipolar excitonic insulator in a moiré lattice},
	volume = {18},
	copyright = {2022 The Author(s), under exclusive licence to Springer Nature Limited},
	issn = {1745-2481},
	url = {https://www.nature.com/articles/s41567-022-01532-z},
	doi = {10.1038/s41567-022-01532-z},
	number = {4},
	urldate = {2025-05-11},
	journal = {Nature Physics},
	author = {Gu, Jie and Ma, Liguo and Liu, Song and Watanabe, Kenji and Taniguchi, Takashi and Hone, James C. and Shan, Jie and Mak, Kin Fai},
	month = apr,
	year = {2022},
	note = {Publisher: Nature Publishing Group},
	pages = {395--400},
}

@article{ma_strongly_2021,
	title = {Strongly correlated excitonic insulator in atomic double layers},
	volume = {598},
	number = {7882},
	urldate = {2025-05-08},
	journal = {Nature},
	author = {Ma, Liguo and Nguyen, Phuong X. and Wang, Zefang and Zeng, Yongxin and Watanabe, Kenji and Taniguchi, Takashi and MacDonald, Allan H. and Mak, Kin Fai and Shan, Jie},
	month = oct,
	year = {2021},
	pages = {585--589},
}

@article{tang_long_2019,
	title = {Long valley lifetime of dark excitons in single-layer {WSe$_2$}},
	volume = {10},
	number = {1},
	urldate = {2025-05-08},
	journal = {Nature Communications},
	author = {Tang, Yanhao and Mak, Kin Fai and Shan, Jie},
	month = sep,
	year = {2019},
	pages = {4047},
}

@article{thureja_electrically_2022,
	title = {Electrically tunable quantum confinement of neutral excitons},
	volume = {606},
	number = {7913},
	urldate = {2025-05-11},
	journal = {Nature},
	author = {Thureja, Deepankur and Imamoglu, Atac and Smoleński, Tomasz and Amelio, Ivan and Popert, Alexander and Chervy, Thibault and Lu, Xiaobo and Liu, Song and Barmak, Katayun and Watanabe, Kenji and Taniguchi, Takashi and Norris, David J. and Kroner, Martin and Murthy, Puneet A.},
	month = jun,
	year = {2022},
	pages = {298--304},
}

@article{shimazaki_strongly_2020,
	title = {Strongly correlated electrons and hybrid excitons in a moiré heterostructure},
	volume = {580},
	number = {7804},
	journal = {Nature},
	author = {Shimazaki, Yuya and Schwartz, Ido and Watanabe, Kenji and Taniguchi, Takashi and Kroner, Martin and Imamoğlu, Ataç},
	month = apr,
	year = {2020},
	pages = {472--477},
}

@article{chen_excitonic_2022,
	title = {Excitonic insulator in a heterojunction moiré superlattice},
	volume = {18},
	number = {10},
	urldate = {2025-05-11},
	journal = {Nature Physics},
	author = {Chen, Dongxue and Lian, Zhen and Huang, Xiong and Su, Ying and Rashetnia, Mina and Ma, Lei and Yan, Li and Blei, Mark and Xiang, Li and Taniguchi, Takashi and Watanabe, Kenji and Tongay, Sefaattin and Smirnov, Dmitry and Wang, Zenghui and Zhang, Chuanwei and Cui, Yong-Tao and Shi, Su-Fei},
	month = oct,
	year = {2022},
	pages = {1171--1176},
}

@article{wang_evidence_2019,
	title = {Evidence of high-temperature exciton condensation in two-dimensional atomic double layers},
	volume = {574},
	number = {7776},
	journal = {Nature},
	author = {Wang, Zefang and Rhodes, Daniel A. and Watanabe, Kenji and Taniguchi, Takashi and Hone, James C. and Shan, Jie and Mak, Kin Fai},
	month = oct,
	year = {2019},
	pages = {76--80},
}

@article{chernikov_exciton_2014,
	title = {Exciton {Binding} {Energy} and {Nonhydrogenic} {Rydberg} {Series} in {Monolayer} {WS$_2$}},
	volume = {113},
	number = {7},
	urldate = {2025-05-11},
	journal = {Physical Review Letters},
	author = {Chernikov, Alexey and Berkelbach, Timothy C. and Hill, Heather M. and Rigosi, Albert and Li, Yilei and Aslan, Burak and Reichman, David R. and Hybertsen, Mark S. and Heinz, Tony F.},
	month = aug,
	year = {2014},
	pages = {076802},
}

@article{deslippe_electronhole_2009,
	title = {Electron-{Hole} {Interaction} in {Carbon} {Nanotubes}: {Novel} {Screening} and {Exciton} {Excitation} {Spectra}},
	volume = {9},
	number = {4},
	urldate = {2025-05-11},
	journal = {Nano Letters},
	author = {Deslippe, Jack and Dipoppa, Mario and Prendergast, David and Moutinho, Marcus V. O. and Capaz, Rodrigo B. and Louie, Steven G.},
	month = apr,
	year = {2009},
	pages = {1330--1334},
	
}

@article{mahan_excitons_1967,
    title = {Excitons in {Degenerate} {Semiconductors}},
    volume = {153},
    number = {3},
    urldate = {2025-05-15},
    journal = {Physical Review},
    author = {Mahan, G. D.},
    month = jan,
    year = {1967},
    pages = {882--889},
}

@article{wakisaka_photoemission_2012,
    title = {Photoemission {Spectroscopy} of {Ta2NiSe5}},
    volume = {25},
    number = {5},
    urldate = {2025-05-15},
    journal = {Journal of Superconductivity and Novel Magnetism},
    author = {Wakisaka, Y. and Sudayama, T. and Takubo, K. and Mizokawa, T. and Saini, N. L. and Arita, M. and Namatame, H. and Taniguchi, M. and Katayama, N. and Nohara, M. and Takagi, H.},
    month = jul,
    year = {2012},
    pages = {1231--1234},
}

@article{chen_anomalous_2023,
	title = {Anomalous excitonic phase diagram in band-gap-tuned {Ta2Ni}({Se},{S})5},
	volume = {14},
	number = {1},
	journal = {Nature Communications},
	author = {Chen, Cheng and Tang, Weichen and Chen, Xiang and Kang, Zhibo and Ding, Shuhan and Scott, Kirsty and Wang, Siqi and Li, Zhenglu and Ruff, Jacob P. C. and Hashimoto, Makoto and Lu, Dong-Hui and Jozwiak, Chris and Bostwick, Aaron and Rotenberg, Eli and Da Silva Neto, Eduardo H. and Birgeneau, Robert J. and Chen, Yulin and Louie, Steven G. and Wang, Yao and He, Yu},
	month = nov,
	year = {2023},
	pages = {7512},
}

@article{ma_multiple_2022,
	title = {Multiple mobile excitons manifested as sidebands in quasi-one-dimensional metallic {TaSe$_3$}},
	volume = {21},
	number = {4},
	urldate = {2025-04-06},
	journal = {Nature Materials},
	author = {Ma, Junzhang and Nie, Simin and Gui, Xin and Naamneh, Muntaser and Jandke, Jasmin and Xi, Chuanying and Zhang, Jinglei and Shang, Tian and Xiong, Yimin and Kapon, Itzik and Kumar, Neeraj and Soh, Yona and Gosálbez-Martínez, Daniel and Yazyev, Oleg V. and Fan, Wenhui and Hübener, Hannes and Giovannini, Umberto De and Plumb, Nicholas Clark and Radovic, Milan and Sentef, Michael Andreas and Xie, Weiwei and Wang, Zhijun and Mudry, Christopher and Müller, Markus and Shi, Ming},
	month = apr,
	year = {2022},
	pages = {423--429},
}

@article{liang_solid_2016,
	title = {Solid state carbon nanotube device for controllable trion electroluminescence emission},
	volume = {8},
	url = {https://pubs.rsc.org/en/content/articlelanding/2016/nr/c5nr07468a},
	doi = {10.1039/C5NR07468A},
	number = {12},
	urldate = {2025-04-23},
	journal = {Nanoscale},
	author = {Liang, Shuang and Ma, Ze and Wei, Nan and Liu, Huaping and Wang, Sheng and Peng, Lian-Mao},
	year = {2016},
	note = {Publisher: Royal Society of Chemistry},
	pages = {6761--6769},
}

@article{volkov_failed_2021,
	title = {Failed excitonic quantum phase transition in {Ta$_2$Ni(Se$_{1-x}$S$_x$)$_5$}},
	volume = {104},
	url = {https://link.aps.org/doi/10.1103/PhysRevB.104.L241103},
	doi = {10.1103/PhysRevB.104.L241103},
	number = {24},
	urldate = {2023-03-19},
	journal = {Physical Review B},
	author = {Volkov, Pavel A. and Ye, Mai and Lohani, Himanshu and Feldman, Irena and Kanigel, Amit and Blumberg, Girsh},
	month = dec,
	year = {2021},
	note = {Publisher: American Physical Society},
	pages = {L241103},
}

@article{li_strong_2017,
    title = {Strong {In}-{Plane} {Anisotropies} of {Optical} and {Electrical} {Response} in {Layered} {Dimetal} {Chalcogenide}},
    volume = {11},
    issn = {1936-0851},
    url = {https://doi.org/10.1021/acsnano.7b04860},
    doi = {10.1021/acsnano.7b04860},
    number = {10},
    urldate = {2025-05-20},
    journal = {ACS Nano},
    author = {Li, Liang and Gong, Penglai and Wang, Weike and Deng, Bei and Pi, Lejing and Yu, Jing and Zhou, Xing and Shi, Xingqiang and Li, Huiqiao and Zhai, Tianyou},
    month = oct,
    year = {2017},
    note = {Publisher: American Chemical Society},
    pages = {10264--10272},
}

@article{pal_pressure-dependent_2024,
    title = {Pressure-dependent excitonic instability and structural phase transition in {Ta$_2$NiS$_5$}: {Raman} and first-principles study},
    volume = {109},
    shorttitle = {Pressure-dependent excitonic instability and structural phase transition in \$\{{\textbackslash}mathrm\{{Ta}\}\}\_\{2\}\{{\textbackslash}mathrm\{{NiS}\}\}\_\{5\}\$},
    url = {https://link.aps.org/doi/10.1103/PhysRevB.109.155202},
    doi = {10.1103/PhysRevB.109.155202},
    number = {15},
    urldate = {2025-05-20},
    journal = {Physical Review B},
    author = {Pal, Sukanya and Sinha, Arijit and Harnagea, Luminita and Telang, Prachi and Muthu, D. V. S. and Waghmare, U. V. and Sood, A. K.},
    month = apr,
    year = {2024},
    note = {Publisher: American Physical Society},
    pages = {155202},
}

@article{jain_commentary_2013,
    title = {Commentary: {The} {Materials} {Project}: {A} materials genome approach to accelerating materials innovation},
    volume = {1},
    issn = {2166-532X},
    shorttitle = {Commentary},
    url = {https://doi.org/10.1063/1.4812323},
    doi = {10.1063/1.4812323},
    number = {1},
    urldate = {2025-06-03},
    journal = {APL Materials},
    author = {Jain, Anubhav and Ong, Shyue Ping and Hautier, Geoffroy and Chen, Wei and Richards, William Davidson and Dacek, Stephen and Cholia, Shreyas and Gunter, Dan and Skinner, David and Ceder, Gerbrand and Persson, Kristin A.},
    month = jul,
    year = {2013},
    pages = {011002},
}

@article{kaneko_orthorhombic--monoclinic_2013,
	title = {Orthorhombic-to-monoclinic phase transition of Ta2NiSe5 induced by the {Bose}-{Einstein} condensation of excitons},
	volume = {87},
	url = {https://link.aps.org/doi/10.1103/PhysRevB.87.035121},
	doi = {10.1103/PhysRevB.87.035121},
	number = {3},
	urldate = {2023-03-27},
	journal = {Physical Review B},
	author = {Kaneko, T. and Toriyama, T. and Konishi, T. and Ohta, Y.},
	month = jan,
	year = {2013},
	note = {Publisher: American Physical Society},
	pages = {035121},
}

@article{mak2013tightly,
  title={Tightly bound trions in monolayer MoS$_2$},
  author={Mak, Kin Fai and He, Keliang and Lee, Changgu and Lee, Gwan Hyoung and Hone, James and Heinz, Tony F and Shan, Jie},
  journal={Nature materials},
  volume={12},
  number={3},
  pages={207--211},
  year={2013},
  }

@article{brown_exciton_1987,
    title = {Exciton binding energy in a quantum-well wire},
    volume = {35},
    copyright = {http://link.aps.org/licenses/aps-default-license},
    issn = {0163-1829},
    url = {https://link.aps.org/doi/10.1103/PhysRevB.35.3009},
    doi = {10.1103/PhysRevB.35.3009},
    number = {6},
    urldate = {2025-05-19},
    journal = {Physical Review B},
    author = {Brown, Jerry W. and Spector, Harold N.},
    month = feb,
    year = {1987},
    pages = {3009--3012},
}

@article{park2018direct,
  title={Direct determination of monolayer {MoS$_2$} and {WSe$_2$} exciton binding energies on insulating and metallic substrates},
  author={Park, Soohyung and Mutz, Niklas and Schultz, Thorsten and Blumstengel, Sylke and Han, Ali and Aljarb, Areej and Li, Lain-Jong and List-Kratochvil, Emil JW and Amsalem, Patrick and Koch, Norbert},
  journal={2D Materials},
  volume={5},
  number={2},
  pages={025003},
  year={2018},
  publisher={IOP Publishing}
}

@article{zhu2015exciton,
  title={Exciton binding energy of monolayer {WS$_2$}},
  author={Zhu, Bairen and Chen, Xi and Cui, Xiaodong},
  journal={Scientific reports},
  volume={5},
  number={1},
  pages={9218},
  year={2015},
  publisher={Nature Publishing Group UK London}
}

@article{maultzsch2005exciton,
  title={Exciton binding energies in carbon nanotubes from two-photon photoluminescence},
  author={Maultzsch, J and Pomraenke, R and Reich, S and Chang, E and Prezzi, D and Ruini, Alice and Molinari, Elisa and Strano, MS and Thomsen, C and Lienau, C},
  journal={Physical Review B},
  volume={72},
  number={24},
  pages={241402},
  year={2005},
  publisher={APS}
}

@article{mohl2018trion,
  title={Trion-polariton formation in single-walled carbon nanotube microcavities},
  author={M\\\"ohl, Charles and Graf, Arko and Berger, Felix J and Luttgens, Jan and Zakharko, Yuriy and Lumsargis, Victoria and Gather, Malte C and Zaumseil, Jana},
  journal={ACS photonics},
  volume={5},
  number={6},
  pages={2074--2080},
  year={2018},
  publisher={ACS Publications}
}

@article{mu2018electronic,
  title={Electronic structures of layered {Ta$_2$NiS$_5$} single crystals revealed by high-resolution angle-resolved photoemission spectroscopy},
  author={Mu, Kejun and Chen, Haiping and Li, Yuliang and Zhang, Yingying and Wang, Pengdong and Zhang, Bo and Liu, Yi and Zhang, GuoBin and Song, Li and Sun, Zhe},
  journal={Journal of Materials Chemistry C},
  volume={6},
  number={15},
  pages={3976--3981},
  year={2018},
  publisher={Royal Society of Chemistry}
}

@article{lampert1958mobile,
  title={Mobile and immobile effective-mass-particle complexes in nonmetallic solids},
  author={Lampert, Murray A},
  journal={Physical Review Letters},
  volume={1},
  number={12},
  pages={450},
  year={1958},
  publisher={APS}
}

@article{wang2018colloquium,
  title={Colloquium: Excitons in atomically thin transition metal dichalcogenides},
  author={Wang, Gang and Chernikov, Alexey and Glazov, Mikhail M and Heinz, Tony F and Marie, Xavier and Amand, Thierry and Urbaszek, Bernhard},
  journal={Reviews of Modern Physics},
  volume={90},
  number={2},
  pages={021001},
  year={2018},
  publisher={APS}
}

@article{ye2021lattice,
  title={Lattice dynamics of the excitonic insulator {Ta$_2$Ni(Se$_{1-x}$S$_x$)$_5$}},
  author={Ye, Mai and Volkov, Pavel A and Lohani, Himanshu and Feldman, Irena and Kim, Minsung and Kanigel, Amit and Blumberg, Girsh},
  journal={Physical Review B},
  volume={104},
  number={4},
  pages={045102},
  year={2021},
  publisher={APS}
}

@article{seki_excitonic_2014,
	title = {Excitonic {Bose}-{Einstein} condensation in {Ta$_2$NiSe$_5$} above room temperature},
	volume = {90},
	url = {https://link.aps.org/doi/10.1103/PhysRevB.90.155116},
	doi = {10.1103/PhysRevB.90.155116},
	number = {15},
	urldate = {2023-03-19},
	journal = {Physical Review B},
	author = {Seki, K. and Wakisaka, Y. and Kaneko, T. and Toriyama, T. and Konishi, T. and Sudayama, T. and Saini, N. L. and Arita, M. and Namatame, H. and Taniguchi, M. and Katayama, N. and Nohara, M. and Takagi, H. and Mizokawa, T. and Ohta, Y.},
	month = oct,
	year = {2014},
	note = {Publisher: American Physical Society},
	pages = {155116},
}

@article{Yale_first,
  title = {Role of electron-phonon coupling in excitonic insulator candidate ${\mathrm{Ta}}_{2}{\mathrm{NiSe}}_{5}$},
  author = {Chen, Cheng and Chen, Xiang and Tang, Weichen and Li, Zhenglu and Wang, Siqi and Ding, Shuhan and Kang, Zhibo and Jozwiak, Chris and Bostwick, Aaron and Rotenberg, Eli and Hashimoto, Makoto and Lu, Donghui and Ruff, Jacob P. C. and Louie, Steven G. and Birgeneau, Robert J. and Chen, Yulin and Wang, Yao and He, Yu},
  journal = {Phys. Rev. Res.},
  volume = {5},
  issue = {4},
  pages = {043089},
  numpages = {16},
  year = {2023},
  month = {Oct},
  }

@article{Rubio_DFT,
	author = {Windg{\"a}tter, Lukas and R{\"o}sner, Malte and Mazza, Giacomo and H{\"u}bener, Hannes and Georges, Antoine and Millis, Andrew J. and Latini, Simone and Rubio, Angel},
	journal = {npj Computational Materials},
	number = {1},
	pages = {210},
	title = {Common microscopic origin of the phase transitions in {Ta$_2$NiS$_5$} and the excitonic insulator candidate {Ta$_2$NiSe$_5$}},
	volume = {7},
	year = {2021},
	}

@article{Dani_dark_exciton,
author = {Julien Madéo  and Michael K. L. Man  and Chakradhar Sahoo  and Marshall Campbell  and Vivek Pareek  and E. Laine Wong  and Abdullah Al-Mahboob  and Nicholas S. Chan  and Arka Karmakar  and Bala Murali Krishna Mariserla  and Xiaoqin Li  and Tony F. Heinz  and Ting Cao  and Keshav M. Dani },
title = {Directly visualizing the momentum-forbidden dark excitons and their dynamics in atomically thin semiconductors},
journal = {Science},
volume = {370},
number = {6521},
pages = {1199-1204},
year = {2020}}

@article{Iliya_review,
author = {Karni, Ouri and Esin, Iliya and Dani, Keshav M.},
title = {Through the Lens of a Momentum Microscope: Viewing Light-Induced Quantum Phenomena in 2D Materials},
journal = {Advanced Materials},
volume = {35},
number = {27},
pages = {2204120},
year = {2023}
}

@article{Exciton_ARPES_WS2,
	author = {Katoch, Jyoti and Ulstrup, S{\o}ren and Koch, Roland J. and Moser, Simon and McCreary, Kathleen M. and Singh, Simranjeet and Xu, Jinsong and Jonker, Berend T. and Kawakami, Roland K. and Bostwick, Aaron and Rotenberg, Eli and Jozwiak, Chris},
	date = {2018/04/01},
	journal = {Nature Physics},
	number = {4},
	pages = {355--359},
	title = {Giant spin-splitting and gap renormalization driven by trions in single-layer {WS$_2$/h-BN} heterostructures},
	volume = {14},
	year = {2018},
	}

@article{SOBOTA,
title = {Ultrafast electron dynamics in the topological insulator {Bi$_2$Se$_3$} studied by time-resolved photoemission spectroscopy},
journal = {Journal of Electron Spectroscopy and Related Phenomena},
volume = {195},
pages = {249-257},
year = {2014},
issn = {0368-2048},
author = {J.A. Sobota and S.-L. Yang and D. Leuenberger and A.F. Kemper and J.G. Analytis and I.R. Fisher and P.S. Kirchmann and T.P. Devereaux and Z.-X. Shen}
}

@article{Bias,
    author = {Gauthier, Nicolas and Sobota, Jonathan A. and Pfau, Heike and Gauthier, Alexandre and Soifer, Hadas and Bachmann, Maja D. and Fisher, Ian R. and Shen, Zhi-Xun and Kirchmann, Patrick S.},
    title = {Expanding the momentum field of view in angle-resolved photoemission systems with hemispherical analyzers},
    journal = {Review of Scientific Instruments},
    volume = {92},
    number = {12},
    pages = {123907},
    year = {2021},
    month = {12},
    issn = {0034-6748},
    doi = {10.1063/5.0053479},
    url = {https://doi.org/10.1063/5.0053479},
}

@article{larkin2017giant,
  title={Giant exciton Fano resonance in quasi-one-dimensional Ta 2 NiSe 5},
  author={Larkin, TI and Yaresko, AN and Pr{\"o}pper, D and Kikoin, KA and Lu, YF and Takayama, T and Mathis, Y-L and Rost, AW and Takagi, H and Keimer, B and others},
  journal={Physical Review B},
  volume={95},
  number={19},
  pages={195144},
  year={2017},
  publisher={APS}
}

@article{windgatter_common_2021,
	title = {Common microscopic origin of the phase transitions in Ta2NiS5 and the excitonic insulator candidate Ta2NiSe5},
	volume = {7},
	copyright = {2021 The Author(s)},
	issn = {2057-3960},
	url = {https://www.nature.com/articles/s41524-021-00675-6},
	doi = {10.1038/s41524-021-00675-6},
	
	number = {1},
	urldate = {2026-01-23},
	journal = {npj Computational Materials},
	publisher = {Nature Publishing Group},
	author = {Windgatter, Lukas and Rosner, Malte and Mazza, Giacomo and Hübener, Hannes and Georges, Antoine and Millis, Andrew J. and Latini, Simone and Rubio, Angel},
	month = dec,
	year = {2021},
	pages = {210},
}

@article{chiba_valence-bond_2019,
	title = {Valence-bond insulator in proximity to excitonic instability},
	volume = {100},
	url = {https://link.aps.org/doi/10.1103/PhysRevB.100.245129},
	doi = {10.1103/PhysRevB.100.245129},
	number = {24},
	urldate = {2026-03-06},
	journal = {Physical Review B},
	publisher = {American Physical Society},
	author = {Chiba, Y. and Mitsuoka, T. and Saini, N. L. and Horiba, K. and Kobayashi, M. and Ono, K. and Kumigashira, H. and Katayama, N. and Sawa, H. and Nohara, M. and Lu, Y. F. and Takagi, H. and Mizokawa, T.},
	month = dec,
	year = {2019},
	pages = {245129},
	}

@article{PhysRevB.94.085111,
  title = {Orbital diamagnetic susceptibility in excitonic condensation phase},
  author = {Sugimoto, Koudai and Ohta, Yukinori},
  journal = {Phys. Rev. B},
  volume = {94},
  issue = {8},
  pages = {085111},
  numpages = {6},
  year = {2016},
  month = {Aug},
  publisher = {American Physical Society},
  doi = {10.1103/PhysRevB.94.085111},
  url = {https://link.aps.org/doi/10.1103/PhysRevB.94.085111}
}

@article{yamada_fflo_2016,
	title = {{FFLO} {Excitonic} {State} in the {Three}-{Chain} {Hubbard} {Model} for {Ta}\$\_2\${NiSe}\$\_5\$},
	volume = {85},
	issn = {0031-9015, 1347-4073},
	url = {http://arxiv.org/abs/1602.03248},
	doi = {10.7566/JPSJ.85.053703},
	number = {5},
	urldate = {2026-01-23},
	journal = {Journal of the Physical Society of Japan},
	author = {Yamada, Takemi and Domon, Kaoru and Ono, Yoshiaki},
	month = may,
	year = {2016},
	note = {arXiv:1602.03248 [cond-mat]},
	pages = {053703},

}

@article{sugimoto_strong_2018,
	title = {Strong {Coupling} {Nature} of the {Excitonic} {Insulator} {State} in \$\{{\textbackslash}mathrm\{{Ta}\}\}\_\{2\}\{{\textbackslash}mathrm\{{NiSe}\}\}\_\{5\}\$},
	volume = {120},
	url = {https://link.aps.org/doi/10.1103/PhysRevLett.120.247602},
	doi = {10.1103/PhysRevLett.120.247602},
	number = {24},
	urldate = {2026-03-12},
	journal = {Physical Review Letters},
	publisher = {American Physical Society},
	author = {Sugimoto, Koudai and Nishimoto, Satoshi and Kaneko, Tatsuya and Ohta, Yukinori},
	month = jun,
	year = {2018},
	pages = {247602},

}

\newpage
\section*{Supplementary materials}

\renewcommand{\thefigure}{S\arabic{figure}}
\renewcommand{\thetable}{S\arabic{table}}
\renewcommand{\theequation}{S\arabic{equation}}
\renewcommand{\thepage}{S\arabic{page}}
\setcounter{figure}{0}
\setcounter{table}{0}
\setcounter{equation}{0}
\setcounter{page}{1}
\subsection{Fitting of the \textit{in-gap} state}

\begin{figure}[h!]
    \centering
    \includegraphics[width=0.5\linewidth]{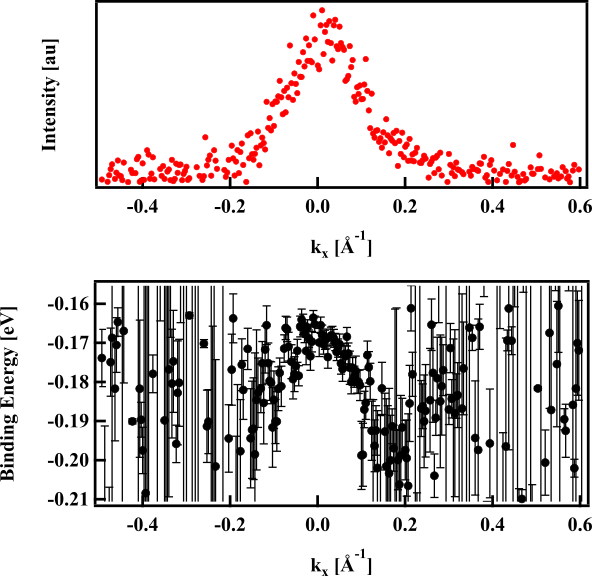}
    \caption{Full momentum range fitting result of the \textit{in-gap} state. Top: peak amplitude as a function of $k_x$. Bottom: peak positions over the full range.}
    \label{fig:sup_ball_fitting}
\end{figure}

Figure \ref{fig:sup_ball_fitting} presents the full momentum range of the \textit{in-gap} state fitting procedure. The EDCs were fitted using Gaussian peak functions. Data was measured with 22 eV photon energy, linear-horizontal polarization and a temperature of 12 K.
The top panel shows the peak amplitudes as a function of $k_x$, highlighting the rapid decrease in the intensity  beyond approximately $\pm0.2$~Å$^{-1}$.
The bottom panel shows the peak positions over the entire range, error-bars illustrating that the fitting does not converge beyond  $\pm0.2$~Å$^{-1}$, the same range shown in the main text. 

\subsection{Out-of-plane ($k_z$) dispersion}
\begin{figure}[h!]
    \centering
    \includegraphics[width=0.5\linewidth]{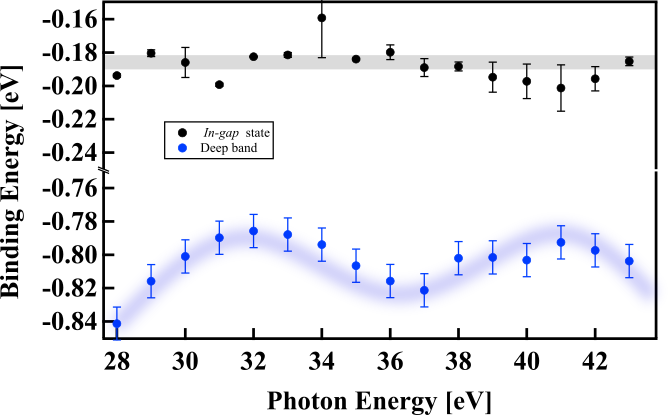}
     \caption{Photon energy dependence of band dispersions in \TNS\ measured by ARPES. The peak positions of a  deep band (blue) and the in-gap state (black) were extracted via Gaussian fitting.  Error bars represent uncertainties from the fitting procedure. Shaded lines are guide-to-the-eye.}
    \label{fig:sup_kz}
\end{figure}

ARPES measurements of \TNS were carried out over a range of photon energies to investigate the out-of-plane momentum dependence ($k_z$) of the electronic structure.
Figure~\ref{fig:sup_kz} presents the $k_z$-dependent dispersions of a deep band and of the in-gap state, extracted by Gaussian fits to the energy distribution curves.
The deeper band (blue) exhibits a clear periodic modulation with photon energy, consistent with dispersion along $k_z$ over approximately two Brillouin zones in the measured photon-energy range.
In contrast, the in-gap state (black) shows no systematic variation with photon energy, indicating a flat dispersion along $k_z$.
Error bars represent the fitting uncertainties.
These results demonstrate that the in-gap state is essentially dispersionless in the out-of-plane direction.

\subsection{Photon polarization dependence}
\begin{figure}[h!]
    \centering
    \includegraphics[width=0.75\linewidth]{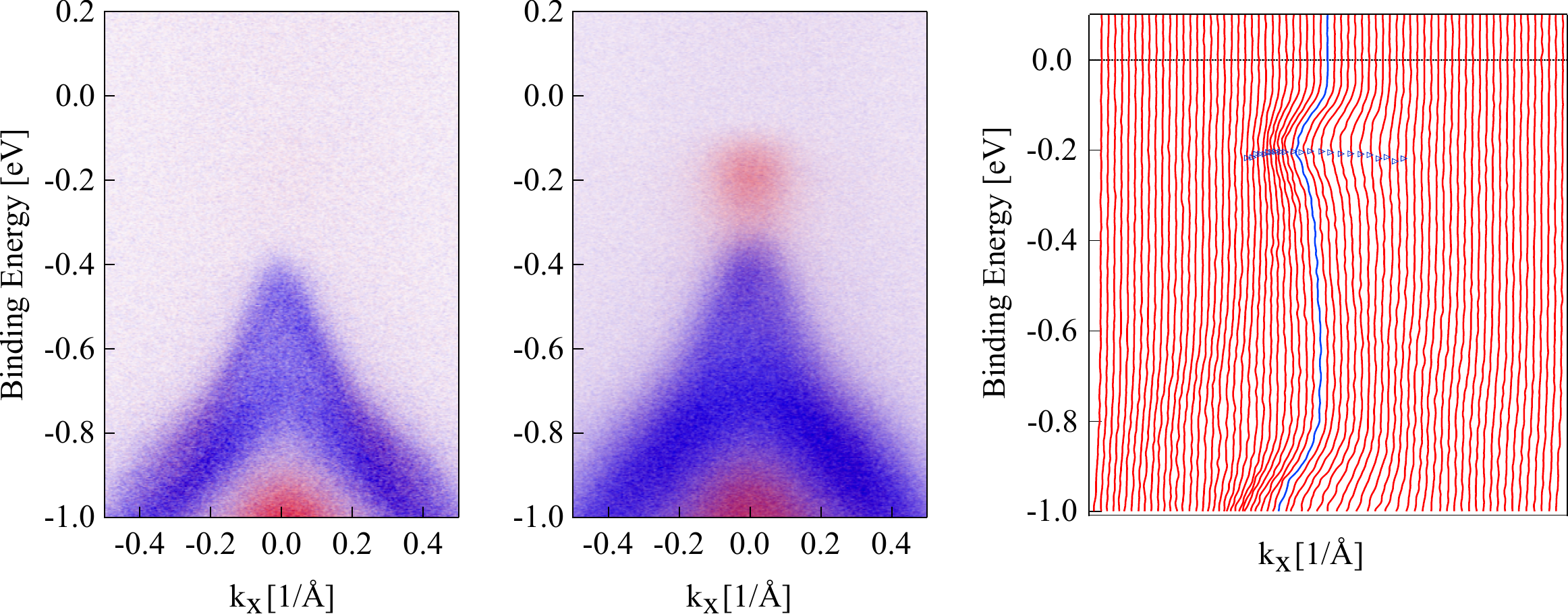}
    \caption{Photon polarization dependence of the valence band and in-gap state. Left and center panels: ARPES spectra measured with vertical (blue) and horizontal (red) photon polarizations for the pristine sample (left) and after 20 s of potassium deposition (center). The valence band shows a polarization dependence consistent with the expected wavefunction symmetry, while the \textit{in-gap} state exhibits the opposite polarization fingerprint, consistent with an electron from a dissociated trion. Right panel show a waterfall plot of the center panel, showing the hole-like dispersion of the \textit{in-gap} state.}
    \label{fig:sup_polarization_dependence}
\end{figure}

Figure \ref{fig:sup_polarization_dependence} illustrates the photon polarization dependence of the valence band and the \textit{in-gap} state. The left and center panels show ARPES spectra acquired with vertical (blue) and horizontal (red) photon polarizations. The leftmost spectrum corresponds to the pristine sample, while the center panel was measured after 20 s of potassium deposition. The valence band exhibits a distinct polarization response, consistent with the expected orbital character \cite{mazza_nature_2020,watson_band_2020}. In contrast, the \textit{in-gap} state displays the opposite polarization fingerprint, consistent with that of the conduction band\cite{mazza_nature_2020,watson_band_2020}, or, as we propose, an electron originating from a dissociated trion. The right panel presents a waterfall plot of the potassium-doped spectrum, highlighting that the \textit{in-gap} feature is indeed gapped showing it cannot be the conduction band. Superimposed peak positions extracted from fits to EDC's reveal the hole-like dispersion of this state (blue line shows k=0). 

\subsection{Temperature dependence}
\begin{figure}
    \centering
    \includegraphics[width=0.5\linewidth]{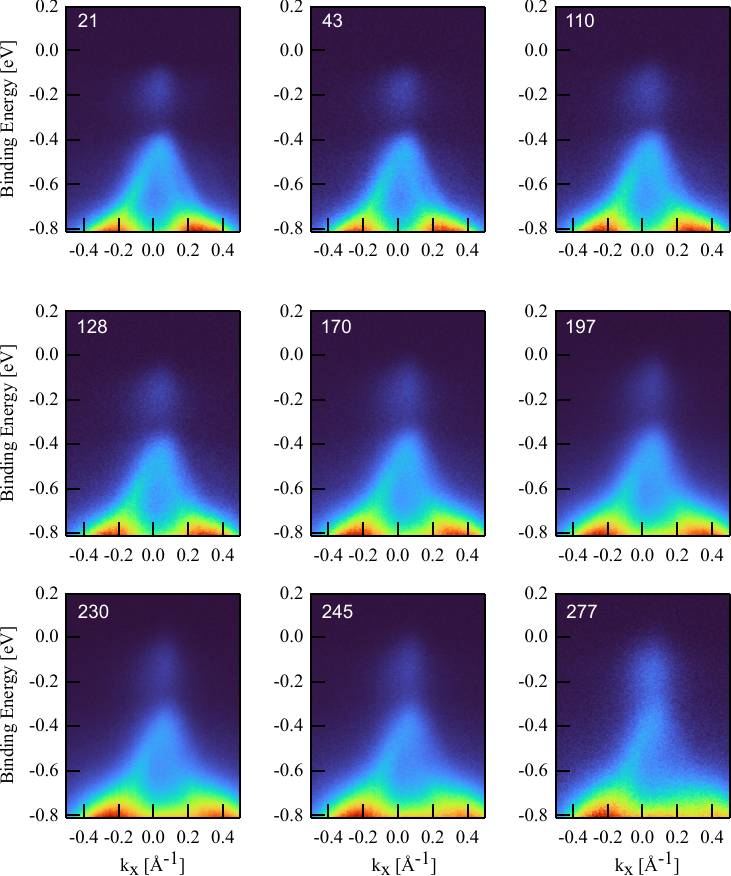}
    \caption{ARPES data along the $\Gamma-X$ direction for temperatures between 21 and 277K}
    \label{fig:temperature_panel}
\end{figure}

\begin{figure}
    \centering
    \includegraphics[width=0.5\linewidth]{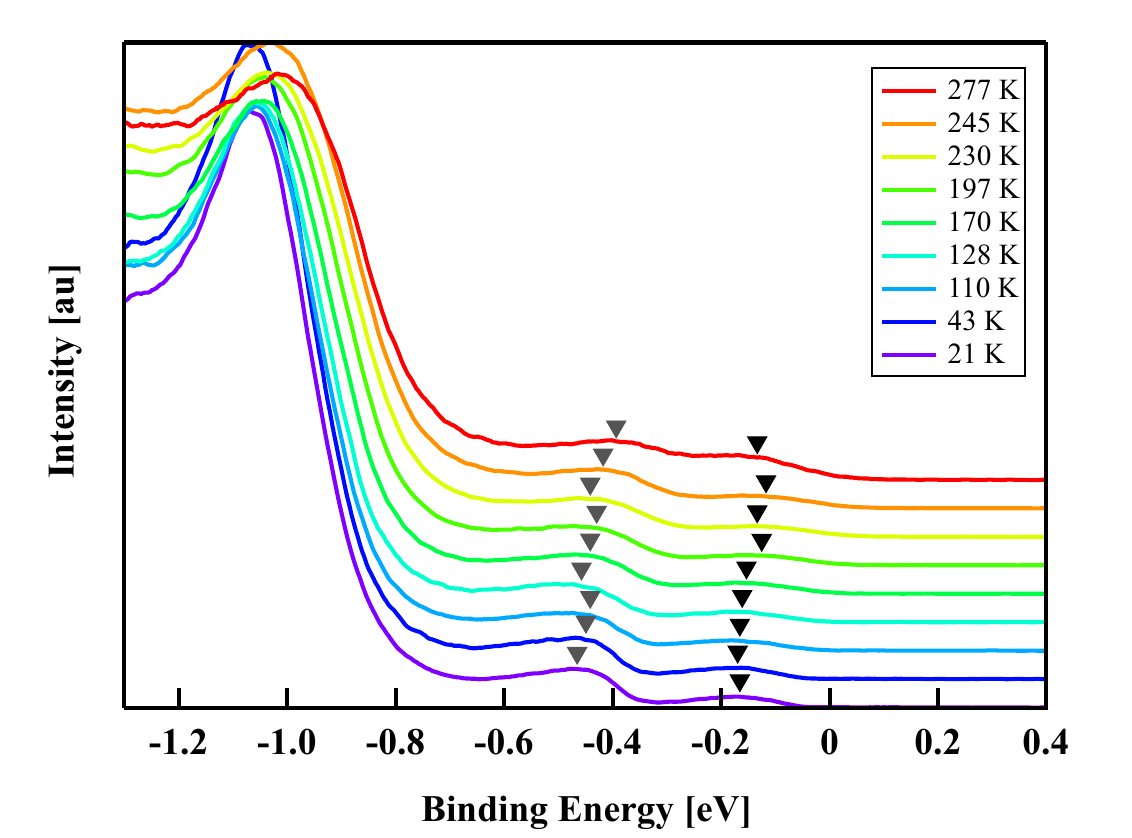}
    \caption{Temperature dependent EDC's at the $\Gamma$ point. Curves are vertically offseted for clarity. Black markers are guides to the eye for the valence band peak and in-gap state.}
    \label{fig:temperature_EDCs}
\end{figure}

Figure \ref{fig:temperature_panel} shows ARPES data along the $\Gamma-X$ direction for temperatures between 21 and 277 K. The in-gap state is clearly visible at all temperatures, although it appears increasingly smeared in energy at elevated temperatures. Fig. \ref{fig:temperature_EDCs} shows EDC's at the $\Gamma$ point extracted from data in Fig. \ref{fig:temperature_panel}.

\subsection{Time evolution at the $\Gamma$ point}
\begin{figure}
    \centering
    \includegraphics[width=0.3\linewidth]{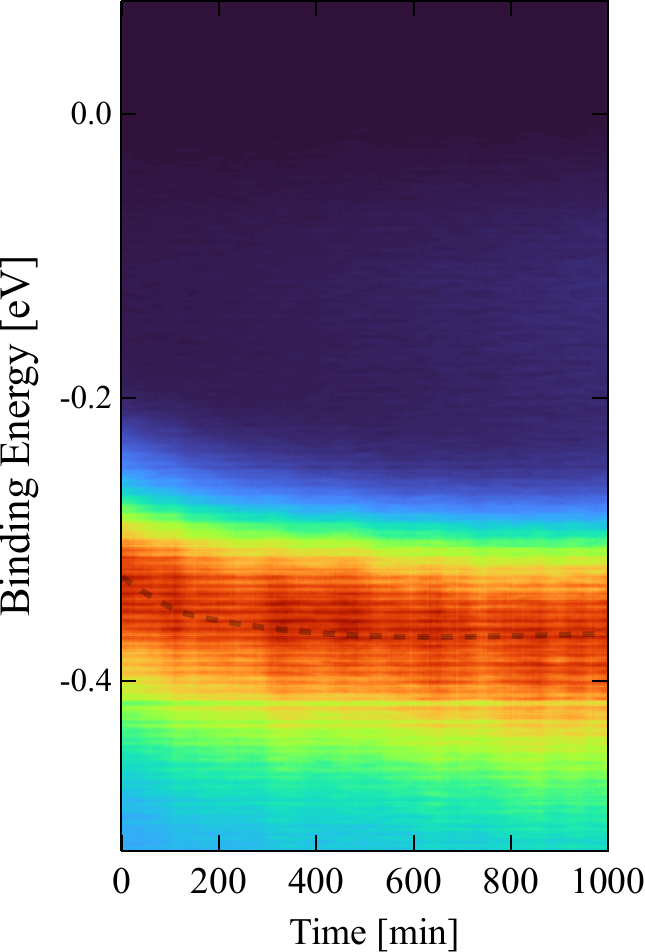}
    \caption{EDCs at the $\Gamma$ point as a function of time after cleave. Dashed line is a guide to the eye for the valence band drift.}
    \label{fig:EDC_time_sup}
\end{figure}

EDCs at the $\Gamma$ point as a function of time after cleaving are presented in Fig.~\ref{fig:EDC_time_sup} . These EDCs were used to extract the \textit{in-gap} state intensity and valence band peak position shown in Fig~\ref{fig:surface_doping_combined} D. The dashed line in Fig.~\ref{fig:EDC_time_sup} serves as a guide-to-the-eye for the valence band drift with time.
\subsection{Conduction band fitting}
\begin{figure}[h]
    \centering
    \includegraphics[width=0.5\linewidth]{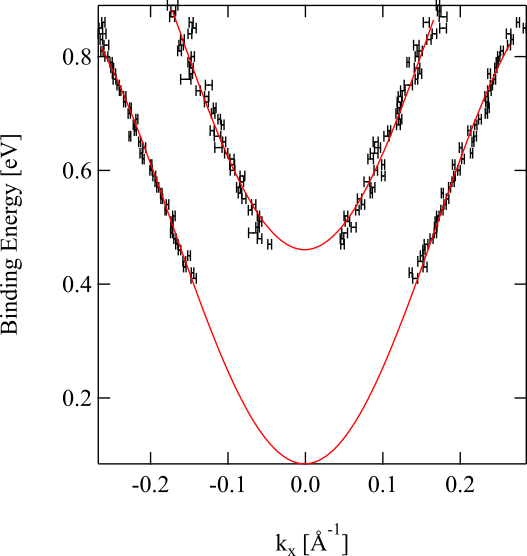}
    \caption{Dispersion fit results for 2PPE data. Black markers show the extracted dispersion from a fit to 4 Lorentzian peaks. Red curves are the results of globally fitting 2 cosine functions to the dispersion, sharing the mass term and the phase}
    \label{fig:sup_fit}
\end{figure}

The conduction band dispersion was extracted using a fit of 4 Lorentzian peaks to the momentum distribution curves (MDCs) of the 2PPE data. The resulting fits are shown in Fig.~\ref{fig:sup_fit} where
we see two parallel bands. Following DFT calculations 
\cite{windgatter_common_2021,sugimoto_strong_2018} we fit the two bands assuming a constant energy separation. The data is fitted with  two cosine functions of the form: \[\epsilon_{c}(k)=y_0+ A\cos(\omega k+\phi)\] forcing a shared mass and phase.

From the fitting results, we determine a conduction band minimum of $\Ec = 84\pm20eV$ and an effective mass of  $0.42\pm0.02 \,\, m_e$.
Although the bottom of the lower conduction band lies outside our data range, the minimum of the inner (upper) band is within range. This, together with the constant shift, improves the reliability of our estimate.  Our model does not account for the k$_z$ dependence, and therefore the absolute minimum of the conduction band may be somewhat lower. 

\subsection{Additional details on the theoretical model} 

\begin{figure}
    \centering

    \includegraphics[width=0.9\textwidth]{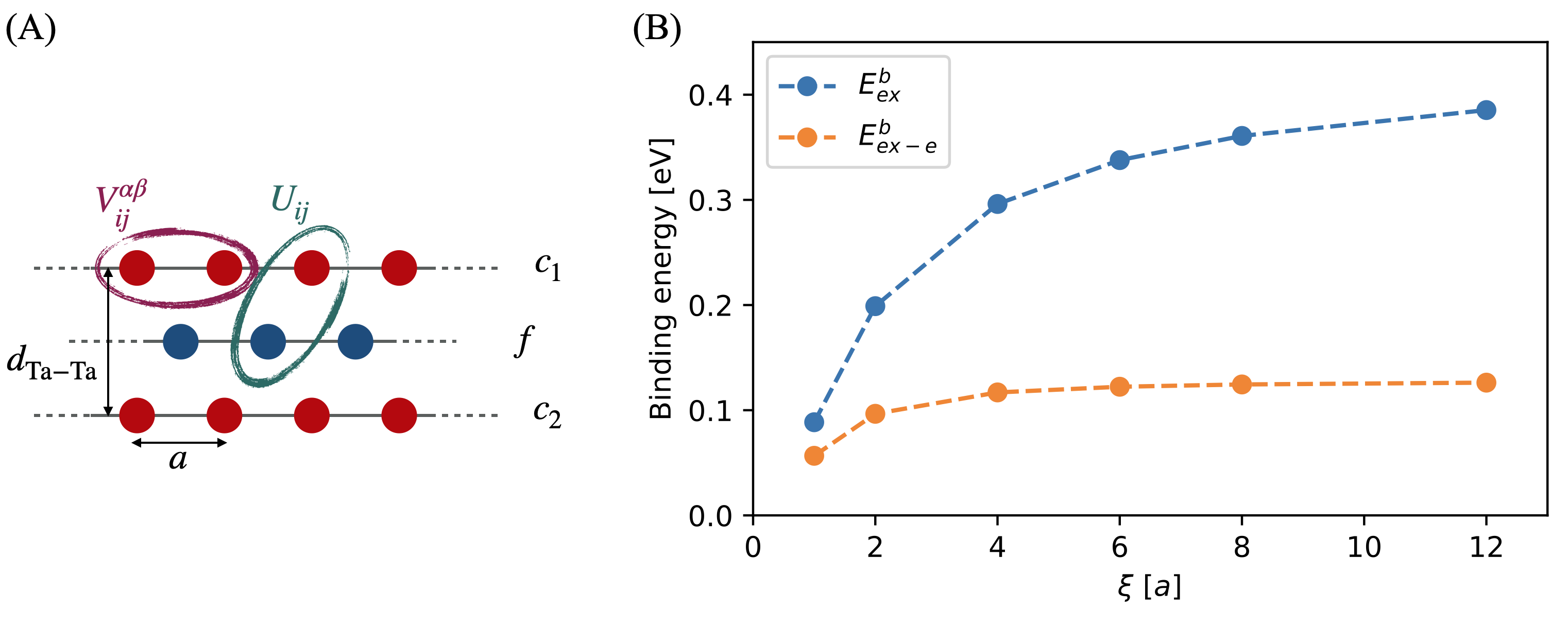}
    \caption{(A) Schematic picture of the one-dimensional model considered. (B) Exciton and exciton-electron binding energies obtained as function of the screening length $\xi$ for model parameters $U_0/a=0.7$eV, $t_c=0.78$eV, and $t_f=1.09$eV.}
    \label{fig:model}
\end{figure}

In this section we present further details on the minimal 1D model, given in Eq.~\eqref{eq:model} in the main text.
A schematic picture of the model is shown in Fig.~\ref{fig:model}(A). 
As mentioned in the main text, the interactions $U_{ij}$, $V^{\alpha\beta}_{ij}$ are modeled using a screened Coulomb potential, namely $U_0e^{-r_{ij}/\xi}/r$, where $U_0=e^2/(4 \pi \epsilon_0 \epsilon_r)$ with $\epsilon_0$ the vacuum permittivity and $\epsilon_r$ the relative permittivity of \TNS, $\xi$ is the screening length and $r_{ij}$ is the respective inter-site distance. 
Using $\epsilon_r=6$ following Ref.~\cite{larkin2017giant} and denoting by $a=3.41$\AA\ the length of the unit cell~\cite{jain_commentary_2013} results in $U_0/a=0.7$eV.
The distance between sites in the conduction and valence band chains is given by $r_{ij} = \left|r_{c,i}-r_{f,j}\right|=\sqrt{(d_{\rm Ta-Ta}/2)^2+a^2(i-j \pm 1/2)^2}$, whereas the distance between sites belonging to the conduction chains is $r_{ij}=\left|r_{c_\alpha,i}-r_{c_\beta,j}\right|=\sqrt{(1-\delta_{\alpha,\beta})\ d_{\rm Ta-Ta}^2+a^2(i-j)^2}$. Here $d_{\rm Ta-Ta}=4.3$\AA\ is the distance between the Ta chains~\cite{jain_commentary_2013}.

To obtain the exciton binding energy we calculate the energy spectrum of the system with a single hole in the valence band and a single electron in the conduction band. A bound state separated from the particle-hole continuum can be clearly observed in the spectrum. Denoting its energy dispersion by $\epsilon_\text{ex}(k)$, the exciton binding energy is given by $E^b_\text{ex}=\epsilon_{\rm ex}(k=0) - \big[ \epsilon_{\rm e}(k=0)+\epsilon_{\rm h}(k=0) \big] $, where $\epsilon_{\rm e(h)}(k)$ is the single particle electron (hole) dispersion in the conduction (valence) band.
The electron-exciton binding energy is obtained similarly, calculating the spectrum within the two electrons and a single hole subspace, where a clear trion bound state can be observed. Denoting its dispersion by $\epsilon_\text{tr}(k)$, we have
$E^b_\text{ex-e}= \epsilon_\text{tr}(k=0) - \big[ \epsilon_{\rm e}(k=0)+\epsilon_{\rm ex}(k=0) \big]$. The trion binding energy, is given by the sum $E^b_\text{tr} = E^b_\text{ex}+E^b_\text{ex-e}$

Figure \ref{fig:model}(B) shows exciton and exciton-electron binding energies as function of the screening length $\xi$, calculated for a system of length $N=40$.
Notably, we find that the exciton-electron binding energy in this system can be significant in comparison to the exciton one.

\newpage

\newpage

\end{document}